\documentclass[fleqn,usenatbib]{mnras}

\usepackage{newtxtext,newtxmath}

\usepackage[T1]{fontenc}
\usepackage{ae,aecompl}
\usepackage{pbox}
\usepackage{xfrac}
\usepackage{mathtools} 
\usepackage{orcidlink}

\usepackage{graphicx}	
\usepackage{amsmath}	
\usepackage{amsfonts}
\usepackage{gensymb}
\usepackage{subfig}
\usepackage{array}
\newcolumntype{P}[1]{>{\centering\arraybackslash}p{#1}}
\newcolumntype{$}{>{\global\let\currentrowstyle\relax}}
\newcolumntype{^}{>{\currentrowstyle}}

\usepackage{todonotes}

\usepackage{etoolbox}
\usepackage{booktabs}
\makeatletter

\makeatletter
\newcommand*{\rom}[1]{\expandafter\@slowromancap\romannumeral #1@}
\makeatother

\makeatother

\title[Transient shielding and BAL/UFO signatures]{Radiation-ionization hydrodynamic simulations of AGN line-driven winds lead to transient shielding and BAL/UFO signatures}

\author[N. Scepi et. al]
{Nicolas Scepi$^{\orcidlink{0000-0003-3909-2486}~1,2}$,
Christian Knigge$^{\orcidlink{0000-0002-1116-2553}~2}$
\thanks{E-mail: c.knigge@soton.ac.uk},
Amin Mosallanezhad$^{\orcidlink{0000-0002-4601-7073}~2}$,
Knox S. Long$^{\orcidlink{0000-0002-4134-864X}~3,4}$,
James H. Matthews$^{\orcidlink{0000-0002-3493-7737}~5}$,
\newauthor{Stuart A. Sim$^{\orcidlink{0000-0002-9774-1192}~6}$ and Austen Wallis$^{\orcidlink{0000-0003-0770-9015}~2}$}
\\
$^{1}$Univ. Grenoble Alpes, CNRS, IPAG, 38000 Grenoble, France\\
$^{2}$School of Physics and Astronomy, University of Southampton, Highfield, Southampton, SO17 1BJ, UK\\
$^{3}$Space Telescope Science Institute, 3700 San Martin Drive, Baltimore, MD, 21218, USA\\
$^{4}$Eureka Scientific Inc., 2542 Delmar Avenue, Suite 100, Oakland, CA, 94602-3017, USA\\
$^{5}$Department of Physics, Astrophysics, University of Oxford, Denys Wilkinson Building, Keble Road, Oxford, OX1 3RH, UK\\
$^{6}$School of Mathematics and Physics, Queen's University Belfast, University Road, Belfast BT7 1NN, UK\\
}

\date{\today}

\pubyear{2023}

\begin{document}
\label{firstpage}
\pagerange{\pageref{firstpage}--\pageref{lastpage}}
\maketitle{}

\begin{abstract}
Disc winds from active galactic nuclei (AGN) can be launched by radiation pressure acting on spectral lines. However, launching a line-driven wind in the X-ray rich environment of AGN is challenging, as the wind easily gets over-ionized. Previous simulations suggested that X-ray self-shielding could enable line driving, though it remained unclear whether this relied on simplified treatments of radiation and ionization. Here, we revisit the X-ray shielding scenario using the first multi-frequency, multi-directional Monte-Carlo radiative photo-ionization hydrodynamical simulations of AGN line-driven winds. We find that sustaining a steady wind with mass-loss rates of $\approx20\%$ of the accretion rate requires an unrealistically weak X-ray flux ($\alpha_{\rm OX}<-3$). For stronger X-ray emission ($-3<\alpha_{\rm OX}<-1$), self-shielding is only transient, leading to episodic ejections with mass-loss rates approaching the accretion rate. Our steady winds naturally produce FeLoBAL, HiBAL, and broad emission line signatures, depending on the disc spectral energy distribution and the observer’s inclination. At moderate X-ray luminosities ($\alpha_{\rm OX}\sim-3$), transient winds can generate short-lived BAL and ultra-fast outflow (UFO) features. At the highest X-ray luminosities ($\alpha_{\rm OX}\sim-1$), the winds are too ionized to form BALs, but still produce UFOs. These results imply that additional physics is required to explain BAL outflows at realistic X-ray levels and to drive winds strong enough for AGN feedback. Nonetheless, our simulations provide a new framework for interpreting the observed diversity of AGN outflow signatures with fully coupled radiation and dynamics.
\end{abstract}



\defcitealias{castor_radiation-driven_1975}{CAK}

\section{Introduction}
\label{section:introduction}

It is widely accepted that active galactic nuclei (AGN) and quasi-stellar objects (QSOs) host winds -- dense, weakly collimated outflows originating from the accretion disc -- that are revealed by the presence of blue-shifted absorption lines in the spectrum. In the UV, Broad Absorption Lines (BAL) are observed in approximately $20\%$ of AGN although this number depends on redshift  \citep{knigge2008,allen2011}. BALs indicate outflowing material with velocities of $0.01-0.1\:c$ \citep{weymann1991,hamann1993,gibson2009,allen2011}, where $c$ is the speed of light. In the X-rays, warm absorbers are associated to a slower wind component  with velocities of $\sim 100 \:\mathrm{km\:s^{-1}}$ \citep{halpern1984,krolik2001} while Ultra-Fast Outflow (UFOs) have the largest velocities inferred so far in AGN winds with velocities up to $\sim 0.3\:c$ \citep{pounds2003,reeves2009,tombesi2010}. 

Winds from QSOs and AGN are thought to play a key role in the co-evolution of black hole and galaxies by providing a source of mass, momentum and  energy to the interstellar and intergalactic medium \citep{ostriker2010,tombesi2012,king2015}. The exact impact of the wind on its environment depends on properties such as its covered solid angle, its kinetic power, its mass-loss rate or its temporal behaviour, providing a strong motivation to better understand the mechanisms behind wind launching in AGN and QSOs.

Several mechanisms have been proposed to explain the formation of AGN disc winds at $L\lesssim L_\mathrm{Edd}$. Thermal launching \citep{begelman1983} occurs when the outer disc gets heated by X-ray irradiation above the local escaped velocity. While it could play a role in driving the wind associated to warm absorbers (\citealt{krolik2001} although see \citealt{mizumoto2019}), it cannot produce outflows as fast as what is inferred for UFOs or BALs. Magnetic launching, driven by a combination of centrifugal force and a magnetic pressure gradient along field lines \citep{blandford1982,ferreira1995}, is a robust and efficient mechanism that can explain the properties of warm absorbers, UFOs, and BALs within a single framework \citep{fukumura2010a,fukumura2010b,fukumura2015}. However, it is generally not favoured in the community due to the lack of observational constraints on the magnetic field structure in accretion discs. The favoured mechanism for wind launching is instead radiation pressure mediated through spectral lines, i.e. line-driven wind \citep{shlosman1985}. Indeed, since this mechanism only requires a relatively high luminosity and the presence of opacity in spectral lines, both of which are observed, it makes it the ideal candidate. Evidence of line-locking also supports line-driving \citep{korista1993,lu2018}.  

Despite the apparent simplicity of line-driven winds, it was soon recognized that launching a line-driven wind in the X-ray rich environment of AGN is a challenge. Indeed, \cite{drew1984} and others carried out simple estimates showing that the ionization level of a wind irradiated by a realistic amount of X-rays should ionize most resonance lines, inhibiting line-driven wind launching.   However, then \cite{murray1995} suggested that X-rays are attenuated in an inner wind propagating in the wind itself allowing efficient line-driving at larger radii. Subsequently, \cite{proga2000} and \cite{proga2004} carried out  2D radiative hydrodynamical simulations supporting this hypothesis. In these simulations, a failed wind from the innermost parts is shielding the X-rays providing conditions that are suitable for line-driving behind it and even for the production of UFO \citep{nomura2016,mizumoto2021} and BAL features \citep{nomura2013}. However, these simulations relies on a quite crude treatment of radiative transfer, assuming that X-rays propagate in straight lines while being attenuated through absorption only. Others, notably \cite{sim2010} and \cite{higginbottom2014} showed that scattering and reprocessing of X-rays could limit the efficiency of shielding as  X-rays would eventually find a way to pass around the shield. Recently, \cite{dyda2024} studied this effect in dynamical simulations by using a multi-directional radiative transfer treatment of the X-rays with ad hoc values for the scattering opacities. They found that in a strongly scattering regime the X-rays would scatter around the shielding wind rendering line-driving inefficient. This reopens the question of whether self-shielding can actually work under a realistic treatment of multi-dimensional radiative transfer taking into account the opacities of the multitude of lines for reprocessing and scattering.

Another limitation of the simulations carried out by \cite{proga2000} and \cite{proga2004} (and subsequent work from others using the same set-up) is the assumption that only X-rays from the central source are responsible for ionizing the wind. However, UV radiation coming from the accretion disc is also able to efficiently ionize the gas and prevent line-driving, as recently shown in accreting white dwarfs \citep{higginbottom2024,mosallanezhad2025}. 

Here we revisit the viability of X-ray shielding in AGN line-driven winds by performing the first multi-frequency, multi-directional radiative transfer and photo-ionization hydrodynamical simulations. With our treatment we are able to realistically evaluate the impact of X-ray scattering and reprocessing as well as X-ray and UV ionization on the line-driving efficiency mechanism in AGN winds. In \autoref{section:method} we describe the technicalities of our simulations, in \autoref{section:results} we present our results, which we then discuss in \autoref{section:discussion} before concluding in \autoref{section:conclusion}.

\section{Method}
\label{section:method}

\subsection{Model Overview}\label{sec:list}
\begin{figure}
    \centering
    \includegraphics[width=80mm]{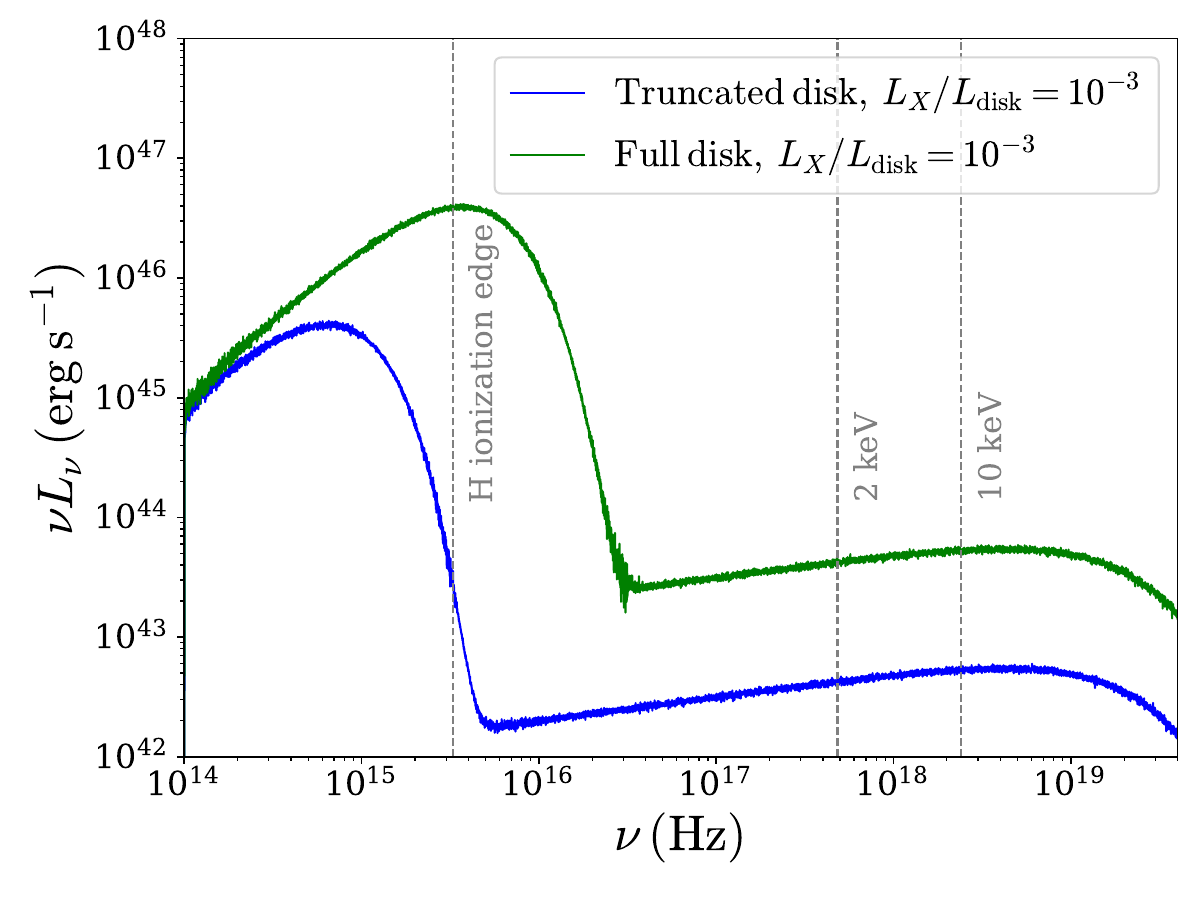}
    \caption{Spectral energy distributions used as input in our simulations.}
    \label{fig:SED_M9}
\end{figure}

We perform a series of ten Monte Carlo radiation-hydrodynamics simulations of line-driven winds in the central parts (between $60$ and $3,000\:r_{\rm g}$) of a $10^9 \:M_\odot$ AGN accreting at a rate of $\dot{M} = 18 \:M_\odot \mathrm{yr}^{-1}$. In these simulations, we explore the impact of different input spectral energy distribution (SED) of the driving radiative flux on the properties of the wind. The parameters of all our simulations are summarized in \autoref{tab:table1}. We use two different types of black-body spectrum to model the disc emission, 
1) a multi-colour blackbody spectrum that peaks at $\approx 8\times10^{14}\:\mathrm{Hz}$, which we refer as a truncated disc SED and 2) a multi-colour blackbody spectrum that peaks at $\approx 4\times10^{15}\:\mathrm{Hz}$, which we refer to as the full disc SED (see \autoref{fig:SED_M9}). We also add a central X-ray bremsstrahlung source (see \autoref{fig:SED_M9}) to model the X-ray emission of AGN. We use five different levels of X-ray emission for each disc SED, probing different regimes of X-ray to disc luminosity, $L_X/L_\mathrm{disc}$, where $L_X$ is the luminosity of the central X-ray source emitted between 2 and 10 keV and $L_\mathrm{disc}$ is the total luminosity of the disc.

The truncated and full disc SEDs differ in at least three ways. First and most importantly, while the full disc SED radiates at almost Eddington ($L_\mathrm{disc}\approx 8.4\times 10^{46}\:\mathrm{erg\:s^{-1}}$), the truncated disc is ten times less luminous radiating at only $L_\mathrm{disc}\approx 7.6\times 10^{45}\:\mathrm{erg\:s^{-1}}$. Secondly, the full disc SED emits two third of its luminosity above the hydrogen ionization energy (13.6~eV), whereas the truncated disc emits only $1\%$ of its luminosity above this energy. Finally, the truncated disc and full disc SEDs produce different illumination geometries relative to the wind domain, with the former emitting most of its energy vertically and the latter emitting most of its energy radially, as we will see in \autoref{sec:sirocco}.

Our choice of disc SEDs is motivated by two observational facts. Many high-mass AGN are observed to lack emission above $\approx2\times 10^{15}$ Hz suggesting that emission from the innermost part of the accretion disc is missing or reprocessed at other frequencies (see \cite{laor2014} and references therein). Our truncated disc SED is supposed to approximately represent this population of AGN. However, there are also a number of AGN with a strong soft X-ray excess that extrapolated to the EUV would emit significantly above $\approx2\times 10^{15}$ Hz (see \cite{kubota2018} and references therein). We approximate this population with our full disc SED although we do not include a realistic model for the soft X-ray excess. 

To assess the impact of a central X-ray source on line-driven AGN winds, we vary the ratio of X-ray to disc emission, $L_X/L_{\rm disc}$, over the wide range $10^{-8}$–$10^{-1}$. This interval is substantially broader than the range inferred from observations. Indeed, estimates of the intrinsic broadband SED of $10^9 M_\odot$ QSOs accreting near the Eddington limit typically yield $L_X/L_\mathrm{disc}\sim10^{-2.5}-10^{-1.8}$ \citep{gallagher2006,mitchell2023}. 
This is in turn quite consistent with estimates of the optical to X-ray spectral index $\alpha_\mathrm{_{OX}}$, defined as 
\begin{equation}\label{eq:alphaOX}
    \alpha_\mathrm{_{OX}}\equiv \log_{10}(\nu L_\nu)_{2\:\mathrm{keV}}-\log_{10}(\nu L_\nu)_{2500\: ,{A}}
\end{equation} 
where $L_\nu$ is the monochromatic luminosity at frequency $\nu$; \cite{temple2023} find $\alpha_\mathrm{_{OX}}\sim-1.8$ to $-1.6$ for luminous quasars. We emphasize that observational estimates of $L_X/L_{\rm disc}$ and $\alpha_{\rm OX}$ are usually derived along a specific line of sight and therefore cannot be directly compared to our input parameter $L_X/L_{\rm disc}$ or to the values of $\alpha_{\rm OX}$ reported in \autoref{sec:obs}. Nevertheless, these measurements provide a useful benchmark for what constitutes a realistic X-ray luminosity, which is sufficient for the purposes of this study.

\subsection{Coupling of PLUTO and SIROCCO}
To perform our radiation-hydrodynamics simulations, we couple two codes -- the hydrodynamical code PLUTO \citep{mignone2009} and the photo-ionization and radiative transfer code SIROCCO \citep{long2002,matthews2025} -- following the procedure described by \cite{higginbottom2024}. Briefly, we use the body force module of PLUTO to include a radiative force. This radiative force is computed from a stand-alone code from \cite{parkin2013} that uses the full ionization state and radiation field in each cell given from SIROCCO to construct force multiplier tables. Crucially, our approach here does not rely on idealised or approximate prescriptions for the force multiplier (see \autoref{sec:force_multiplier}).
These force multiplier tables are then used in PLUTO to compute the radiative force at each hydrodynamical time step, $\Delta t_{\rm HD}$. Because every call to SIROCCO is expensive, we only update the radiative field and the ionization states of the wind every $\Delta t_{\rm RAD}$, where $\Delta t_{\rm RAD}\gg \Delta t_{\rm HD}$. We start our PLUTO-SIROCCO simulation by calling SIROCCO every $1/(2\Omega_\mathrm{in})$, where $\Omega_\mathrm{in}$ is the Keplerian angular velocity at the inner radius. Once the outer parts of the wind have started to develop, we call SIROCCO every $5/\Omega_\mathrm{in}$ with the aim of allowing an approximately steady-state wind to form in the outer parts of the simulation domain in a reasonable computing time. This corresponds to $\Delta t_{\rm RAD}\approx 6000 \Delta t_{\rm HD}$. Sometimes, during rapid dynamic changes (see \autoref{sec:intermediate}), our method might require smaller $\Delta t_{\rm RAD}$. When this is the case, we have not found large differences in the wind behaviour due to the reduced $\Delta t_{\rm RAD}$. The coupling between the two codes has been tested and used extensively in the case of CVs \citep{higginbottom2024,mosallanezhad2025,mosallanezhad2026}. However, we have made several adjustments in our PLUTO and SIROCCO setup for the conditions specific to AGN that we will detail below. 
\newline

\subsection{PLUTO specifics}\label{sec:pluto}

We solve the hydrodynamics (HD) equations with a HLL solver, a piecewise linear (Van Leer) method reconstruction scheme and a second-order Runke-Kutta time integrator. We perform 2.5D simulations on a spherical polar grid going from $60\:r_g$\footnote{As we emphasize later, the inner edge of the grid in PLUTO and SIROCCO can be different so that we can take into account radiation coming from radii below $60\:r_{\rm g}$.} to $3000\:r_{\rm g} $ (i.e. $8.8\times10^{15}\:\mathrm{cm}$ to $4.4\times10^{17}\:\mathrm{cm}$ for a black hole mass of $M_\mathrm{BH}=10^9 M_\odot$) in the radial direction and from $0$ to $\pi/2$ in $\theta$. We use a logarithmic grid with $100$ cells in $r$ and a stretched grid with $140$ cells in $\theta$. We use a stretching coefficient of $0.938$ so as to concentrate the resolution near the midplane of the disc and resolve the sonic point \citep{proga1998}. We use a purely outflowing boundary condition at the inner and outer radial boundaries and an outflowing condition taking care of the axis at the inner latitudinal boundary. Finally, at the midplane we enforce a Keplerian velocity for the azimuthal velocity and a density going as $\rho_{\rm _{BC}}(R) = \rho_{_0} (R_{_0}/R)^{-1.5} $ where $R$ is the cylindrical radius, $ R_{_0}=60\:r_{\rm g} $ and $\rho_{_0} = 10^{-10}\:\mathrm{g\:cm^{-1}}$. We note that the disc is not evolving, i.e. there is no accretion as we do not put any viscosity in the disc. Finally, in order to treat the funnel near the pole, we set a low density floor at $10^{-24}\:\mathrm{g\:cm^{-1}}$. We also put a cap on the velocity at the speed of light. This cap is reached only in the empty funnel, whose dynamics is unphysical because of the use of density floors. The funnel does not impact the main flow dynamic so that the use of this cap is not determinant for our conclusions.

A challenge of AGN wind simulations in a disc system is to resolve the sonic point, given the low ratio of sound speed, $ c_{s} $, to Keplerian velocity, $ v_{\rm _K} $, in AGN, which is $c_s/v_{\rm _K} \lesssim10^{-2} $. The traditional way of resolving the sonic point is to chose a density boundary condition $ \rho_{_0} $ that is high enough to reach regions of the disc that are below the sonic point. We find that this approach can be limited within the context of Monte-Carlo simulation in the very high luminosity regime of AGN. Indeed, in theory the radiative force should be small in the disc's optically thick bulk and should gradually reach the disc input radiative flux above the photosphere. This should ensure that the base of the disc is gradually accelerated, reaching the sonic point after a few cells in $\theta$. However, we find that, in practice, Monte-Carlo noise in the optically thick base of the wind can drive a very strong wind that becomes supersonic right from the first cell of the domain.

To alleviate this issue, we manually ensure in PLUTO that the radiative force cancels in the optically thick regions, $t_i\gg1$, where
\begin{equation}
t_i \equiv \sigma_\mathrm{e} \rho \, v_\mathrm{th}
\left|\frac{\mathrm{d}\!\left(\vec{v}\cdot \hat{n}_i\right)}{\mathrm{d}s_i}\right|^{-1}
\end{equation}
is the Sobolev optical depth in the direction $ i $, $ |\mathrm{d}(\vec{v}\cdot \hat{n}_i)/ \mathrm{d}s_i|$ the gradient of velocity along a given direction $i$, $\rho$ the density, $v_\mathrm{th}$ the thermal velocity of the gas and $ \sigma_\mathrm{e} $ the Thomson scattering opacity. More precisely, we multiply the radiative force in each direction $i$ (we used 36 angles) by $0.5\times(1-\tanh[100\times \log_{10}(t_i)])$. With this extra factor, we are able to resolve the sonic point by typically 10 to 20 points. We made sure that our results are not sensitive to the choice of $ \rho_{_0} $ provided that it sets a density that is deep enough in the disc to produce an optically thick base.

In order to improve the resolution of the sonic point, we also used a ``trick'' when dealing with the temperature in PLUTO. In the hydrodynamic solver of PLUTO, we use an isothermal equation of state with a constant temperature of $3\times10^4$ K so that the sonic point gets to higher $z/x$ (see \autoref{fig:force_multiplier}) at larger radii. However, when computing the Sobolev optical depth and when passing the temperature to SIROCCO, we use a temperature that is isothermal on cylinders and going as $T=(3 G M_\mathrm{BH}\dot{M}/8\pi\sigma R^3)^{1/4}$ as expected from a thin standard disc \citep{Shakura}. This procedure avoids a potential inconsistency with the temperature of the disc in SIROCCO, which is a multi-color blackbody. We tested the dependence of our results by restarting a simulation with a temperature that is constant in cylinders also in the hydrodynamics of PLUTO and do not find that it leads to significant differences in the result. Moreover, we computed a posteriori the temperature from SIROCCO (see \autoref{appendix:isothermal}) and find that it is not far from $3\times10^4$ K in the regions of interest, so that our isothermal assumption seems to be reasonable as a first approximation.

\subsection{SIROCCO specifics}\label{sec:sirocco}

At the beginning of each SIROCCO call we import the velocity, density and temperature fields \footnote{although electron temperature is fixed and is isothermal in cylinders as explained in \autoref{sec:pluto}} from PLUTO. During each SIROCCO iteration, we perform three ionization cycles with the fixed temperature mode using $10^7$ photons. We start each SIROCCO iteration using the ionization structure from the previous iteration as an initial condition. 

To set the two types of SEDs described in \autoref{sec:list}, we use the built-in spectral models of SIROCCO. For the truncated disc SED models we use a multi-colour blackbody disc extending to an inner radius of $r_\mathrm{in,\:SIROCCO}=60\:r_g$, while for the full disc SED we use $r_\mathrm{in,\:SIROCCO}=6\:r_g$. We emphasize that while the inner radius of the disc in PLUTO, $r_\mathrm{in,\:PLUTO}$, is always set at $60\:r_g$, the inner radius of the disc in SIROCCO, $r_\mathrm{in,\:SIROCCO}$, can be smaller than 60 $r_g$ since the two codes run separately and can use different grids. When $r_\mathrm{in,\:SIROCCO}<r_\mathrm{in,\:PLUTO}$, radiation evolves freely in the regions where $r_\mathrm{\:SIROCCO} < r_\mathrm{in,\:PLUTO}$ and interacts with gas only when entering the PLUTO domain. As such, we can take into account radiation from a disc covering a larger range of radii than what we dynamically evolve in PLUTO. When $r_\mathrm{in,\:SIROCCO}\ll r_\mathrm{in,\:PLUTO}$, most of the disc radiation is effectively coming from a point source located at the centre of the spherical PLUTO domain. This is why the truncated and full disc SEDs have different radiation field geometries, the former emitting most of its energy vertically and the latter radially. For the X-rays, we use a central source with a bremmsstralhung spectrum going as $L_\nu\propto\nu^{\alpha}e^{-h\nu/k_BT}$, with $\alpha=-0.8$ and $T=10^9$ K. The intensity of this central X-ray source is set by our choice of $L_X/L_\mathrm{disc}$ (see \autoref{tab:table1}).

\begin{table*}
    \centering
    \caption{Summary of the parameters and some results on dynamical properties of our simulations. The X-ray levels (low, intermediate and high) are used to describe the wind dynamics in \autoref{sec:structure} and refer broadly to whether the X-rays affect the disc dynamics weakly, significantly, or strongly.}
    \setlength\tabcolsep{0pt}
    \begin{tabular*}{\linewidth}{@{\extracolsep{\fill}} ccccccc }
        \midrule
        \midrule
        $L_X/L_\mathrm{disc}$ & X-ray level & \textbf{Status of wind} & \textbf{Average mass-loss rate} & \textbf{Peak mass-loss rate} & \textbf{Average kinetic luminosity} & \textbf{Peak kinetic luminosity} \\
         & & ($M_\odot\:\mathrm{yr}^{-1}$) & ($M_\odot\:\mathrm{yr}^{-1}$) & ($L_\mathrm{Edd}$) & ($L_\mathrm{Edd}$) \\

        \midrule
        \midrule
            \multicolumn{7}{c}{\bfseries{Full disc SED}} \\
        \midrule
         $10^{-6}$ & ``Low" & Strong & 3.3 & 7.4 & $2.9\times 10^{-2}$ & $1.0\times10^{-1}$  \\  
          $10^{-4}$ & ``Intermediate" & Variable & $9.5\times 10^{-1}$ & $6.1\times10^1$ & $1.7\times 10^{-3}$ & $1.7\times10^{-1}$ \\
         $10^{-3}$ & ``Intermediate" & Variable & $1.7$ & $3.0\times10^{1}$ & $4.9\times 10^{-3}$ & $2.9\times10^{-1}$  \\
         $10^{-2}$ & ``High" & Maybe variable & $1.0\times 10^{-1}$ & 7.0 & $3.6\times10^{-4}$ & $4.2\times10^{-2}$\\  
         $10^{-1}$ & ``High" & Variable & $2.7\times10^{-1}$ & $6.7$ & $4.5\times10^{-4}$ & $8.0\times10^{-2}$ \\  
        \midrule
         \multicolumn{7}{c}{\bfseries{Truncated disc SED}} \\
        \midrule
         $10^{-8}$ & ``Low" & Steady & 2.4 & $1.4\times10^1$ & $4.3\times 10^{-3}$ & $3.2\times10^{-1}$  \\  
         $10^{-6}$ & ``Low" & Steady & 1.7 & 8.0 & $1.6\times 10^{-3}$ & $3.3\times 10^{-2}$  \\ 
          $10^{-5}$ & ``Intermediate" & Steady & $5.6\times 10^{-1}$ & 1.8 & $7.5\times 10^{-5}$ & $1.9\times 10^{-3}$  \\ 
          $10^{-4}$ & ``Intermediate" & Variable & $4.9\times 10^{-2}$ & $1.1\times10^1$ & $1.4\times 10^{-3}$ & 1.5 \\ 
          $10^{-3}$ & ``High" & Suppressed & $1.9\times 10^{-5}$ & $1.1\times 10^{-2}$ & $2.2\times 10^{-6}$ & $9.0\times 10^{-4}$  \\ 
        \midrule
    \end{tabular*}
    \label{tab:table1}
\end{table*}

Because we use a disc density that is quite high in order to resolve the sonic point, the atmosphere of the disc redirects (through scattering or reprocessing) a lot of the input disc photons back to the disc boundary condition. Our choice of disc boundary needs to be able to deal with these back-scattering photons to be sure that we indeed input the desired SED. We tried two energy-conserving modes: 1) the disc boundary reflects the photons so that they are kept in the simulation, 2) the disc boundary absorbs the photons that are then used to heat the disc increasing its effective temperature for the next ionization cycle, what we call backwarming in \cite{mosallanezhad2026}. This can be expressed as 
\begin{equation}
(T_\mathrm{eff}^{n+1})^4 = T_\mathrm{eff,SS}^4+r^n(T_\mathrm{eff}^{n})^4
\end{equation}
where $T_\mathrm{eff,SS}$ is the standard Shakura-Sunyaev \citep{Shakura} solution and $T_\mathrm{eff}^{n}$ is the effective temperature of the disc for the ionization cycle $n$. $r^n$ is the coefficient of reflection of the disc's atmosphere, which can be different for each ionization cycle. In the most simple case, where $r$ is independent of the ionization cycle, we expect this sequence to converge to $T_\mathrm{eff,SS}^4/(1-r)$. We verified that in a more complicated case, our sequence does converge in $\approx 20$ cycles although this number depends on $\rho_{_0}$. At each call to SIROCCO, we only perform 3 iterations to compute the ionization. However, we have verified that after several calls to SIROCCO (with PLUTO calls in between) our sequence also converges. 

To mitigate the impact of using high disc densities, we cut out all cells that have a maximum optical depth to Thomson electron scattering (within a cell) larger than 30 across their radial extent (their largest one). Excising these cells speeds up the calculation by avoiding the need to deal with extremely optically thick gas, which is less reliably modelled in SIROCCO anyway. We made sure that this procedure does not impact our results by re-running our simulations for a fraction of time with a higher threshold on the maximum optical depth. 

Finally, to be sure to conserve energy at each matter/photon interaction within our dense disc atmosphere or wind, we used the hybrid macro-atom scheme of SIROCCO \citep{matthews2025}, which enforces strict co-moving frame energy conservation at the interaction point for each Monte Carlo photon packet.

Another consequence of the extreme luminosity of AGN is that the efficiency of radiative driving in the funnel, which depends on the chosen density floor, can cause numerical artefacts if one is not careful. Indeed, we find that we had to add an X-ray source, even very weak, to ionize this region and ensure that the force multiplier is effectively zero. Otherwise, even small values of force multipliers can produce such large forces at the inner boundary of the simulation that they create an artificial injection of mass in the domain. In principle, this artificial mass injection could also be avoided by setting the Riemann flux to zero at the inner funnel boundary; however, this is not a trivial thing to do in PLUTO and since we avoided this problem by systematically adding an X-ray source (again even very weak) such a modification was not needed. 

\subsection{Force multiplier tables and radiative force}
\label{sec:force_multiplier}
The link between PLUTO and SIROCCO is made through the radiative force defined as 
\begin{equation}
\vec{g}_\mathrm{rad}=\sum_i(1+\mathcal{M}(t_i))\sigma_e \frac{\vec{F}_{\mathrm{UV},i}}{c}
\end{equation}
where $\vec{F}_{\mathrm{UV},i}$ is the directional UV flux, i.e. integrated over frequencies between $7.4\times10^{14}$ Hz and $3\times10^{16}$ Hz. 
The directional radiative UV flux comes from SIROCCO. We stress here that we do not include driving by the X-ray flux. This could potentially lead to further driving as noticed by \cite{dannen2019,dyda2025}. We will discuss this point further in \autoref{sec:comparison}.

To compute $\mathcal{M}(t_i)$ we interpolate from a force multiplier table that is created by a stand alone code that possesses a larger line list than that of SIROCCO, containing over 450,000 lines (see \citealt{parkin2013}). This code takes as entry the ionization structure of the plasma and the mean radiative intensity, $J_\nu$, to compute 
\begin{equation}\label{eq:method_force_multiplier}
\mathcal{M}(t) = \sum_\mathrm{lines} \Delta\nu_D \frac{J_\nu}{J}\frac{1-\exp(-\tau^S_{u,l})}{t}
\end{equation}
where $\Delta\nu$ is the Doppler width of a line and $\tau^S_{u,l}$ is the Sololev optical depth of a spectral line between a lower state $l$ and an upper state $u$ defined as in \cite{parkin2013}. We note that we add an extra cap on the force multiplier at $\mathcal{M}_\mathrm{max}=4400$ to avoid extreme force multipliers in the optically thin limit. 

\begin{figure*}
    \centering
    \includegraphics[width=\textwidth]{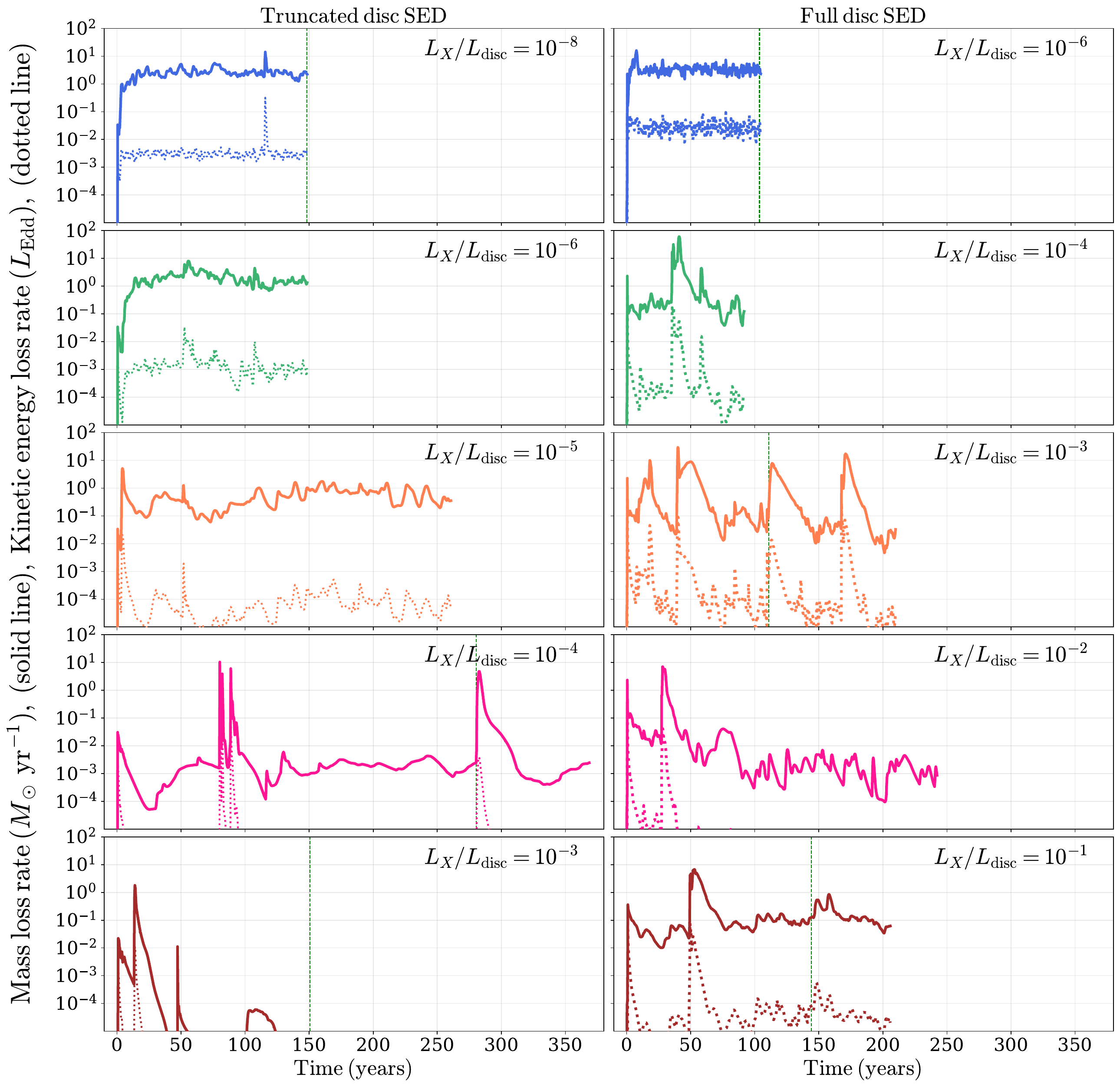}
    \caption{Mass-loss rate in solar mass per year (solid lines) and kinetic luminosity in Eddington units (dotted lines) as a function of time in years for all of our simulations. The left column shows the simulations with the truncated disc SED while the right column shows the simulations with the full disc SED. From top to bottom the X-ray luminosity of the central source increases. The green lines show the times at which we make snapshots and spectra in the rest of the paper. }
    \label{fig:massloss}
\end{figure*}

\section{Results}
\label{section:results}

For better clarity in the presentation of our results, we classify the simulations under the broad categories of low, intermediate, and high X-ray levels, depending on whether the X-rays affect the disc dynamics weakly, significantly, or strongly. The designation of each run is listed in \autoref{tab:table1}.
 
\subsection{Mass-loss and kinetic luminosities}\label{sec:massloss}

We plot on \autoref{fig:massloss}, for all of our simulations, the mass-loss rate and the kinetic luminosity, defined respectively as
\begin{equation}
\dot{M}_w=\int_0^{\pi/2}\dot{m}_w d\theta=4\pi r^2\int_0^{\pi/2}\rho \max (v_r,0)\sin(\theta)d\theta 
\end{equation}
and 
\begin{equation}
\dot{E}_{K,w}=\int_0^{\pi/2}\dot{e}_{K,w} d\theta=4\pi r^2\int_0^{\pi/2} \frac{1}{2} \rho v^2 \max (v_r,0)\sin(\theta)d\theta,
\end{equation}
where $v$ is the magnitude of the total velocity. We also show in \autoref{tab:table1} the values of the time-averaged and peak mass loss rates and kinetic luminosities for each simulation. We start the time average after 20 years. 

\begin{figure*}
    \centering
    \includegraphics[width=\textwidth]{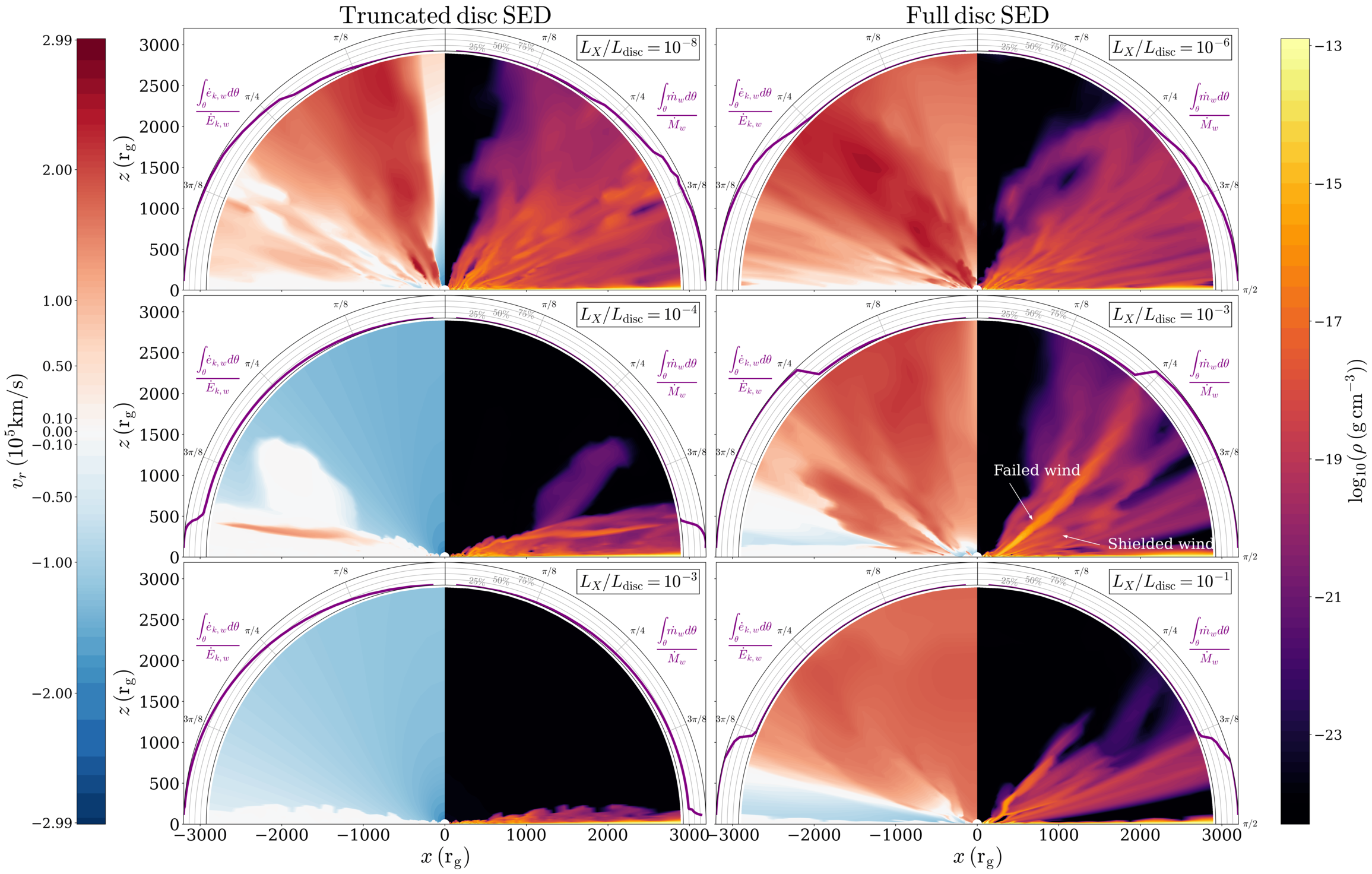}
    \caption{Density (right) and velocity (left) maps of 6 selected simulations. The left column shows the simulations with the truncated disc SED while the right column shows the simulations with the full disc SED. From top to bottom the X-ray luminosity of the central source increases. On each colorplot a polar plot shows the cumulative mass loss rate normalized by the total mass loss rate (on the density colormap) and the cumulative kinetic luminosity normalized by the total kinetic luminosity (on the velocity colormap) to give a sense on where mass and kinetic energy are lost.}
    \label{fig:density_velocity}
\end{figure*}

For the lowest X-ray levels, the mass-loss rate and kinetic luminosity become steady after a few tens of years (less than a hundred orbital periods at the inner edge of the disc). The mass-loss rate stabilizes at a time-averaged value of $\approx 2.4\:M_\odot\:\mathrm{yr}^{-1}$ for the truncated disc SED and $\approx 3.3\:M_\odot\:\mathrm{yr}^{-1}$ for the full disc SED. The fact that these rates are comparable is, at first, surprising; the luminosity in the full disc case is ten times higher than in the truncated case, so one might expect a correspondingly stronger outflow. As we will see in \autoref{sec:structure}, the comparable mass-loss rates reflect the over-ionization of the wind by the disc continuum that peaks at higher energy in the full disc case compared to the truncated case. The kinetic power stabilizes at a time-averaged value of $\approx 8.5\times10^{-3}\:L_\mathrm{Edd}$ for the truncated disc SED and at $\approx 5.8\times10^{-2}\:L_\mathrm{Edd}$ for the full disc SED. It is notable that the kinetic luminosity is ten times higher in the full disc SED compared to the truncated disc SED when the mass-loss rate is of the same order for both cases. This means that the increase in the kinetic power rate is mostly due to an increase in the velocity of the wind when going from the truncated to the full disc SED.

For the truncated disc SED, as we increase the X-ray level up to $L_X/L_\mathrm{disc}=10^{-5}$, the wind stays steady and the time-averaged mass-loss decreases. However, as we reach $L_X/L_\mathrm{disc}=10^{-4}$, for both the truncated disc and full disc SED, the ejections become transient with the disc alternating between periods of low mass loss and sudden bursts of ejections where the mass-loss rates and kinetic luminosities can increase by as much as 3 orders of magnitude. These bursts of ejections follow a fast rise, exponential decay behaviour. Indeed, they are very sudden happening on a year timescale and they decay on a several tens of years timescale, reflecting the time for the transient outflow to cross the outer boundary. It is unclear whether or not this bursty behaviour is quasi-periodic, as we cannot run the simulations for a long enough time to obtain sufficient statistics. 

Finally, for the highest X-ray levels, the outcome depends on our choice of disc SED. For the truncated disc SED, the wind completely dies out. However, for the full disc SED, the wind is never completely suppressed. In fact, the wind gets stronger for $L_X/L_\mathrm{disc}=10^{-1}$ than for $L_\mathrm{disc}/L_x=10^{-2}$. This enhancement in the wind strength is simply due to the contribution to the UV flux of the low frequency part of the central X-ray power-law source, which increases the driving flux by roughly $10\%$. We do not run a simulation for a larger $L_X/L_\mathrm{disc}$ as we do not include driving in X-ray lines (see \autoref{sec:comparison}). In any case, we see that for $L_X/L_\mathrm{disc}=10^{-1}$ and the full disc SED, there are peaks in the mass-loss rate where it can reach values as high as $6.7\:M_\odot\:\mathrm{yr}^{-1}$, a third of the accretion rate.

As can be seen from \autoref{fig:massloss}, our choice of disc SED is crucial. For the truncated disc SED, the wind is almost completely suppressed for $L_X/L_\mathrm{disc}=10^{-4}$, which is a relatively low X-ray level. Contrarily, for the full disc SED, the wind is never totally suppressed. We believe that this is due to two effects. First and most importantly, the disc is two orders of magnitude less luminous in the truncated case than in the full disc case so that the launching of the wind really relies on the presence of spectral lines. Secondly, in the full disc SED the driving and ionizing radiation are both aligned (mostly radial). This configuration is better suited to shielding (see \autoref{sec:structure}) than the case of a truncated disc SED where the driving and ionizing radiation are almost perpendicular at the base of the wind.

\subsection{Dynamical structure of the wind}\label{sec:structure}

\begin{figure}
    \centering
    \includegraphics[width=90mm]{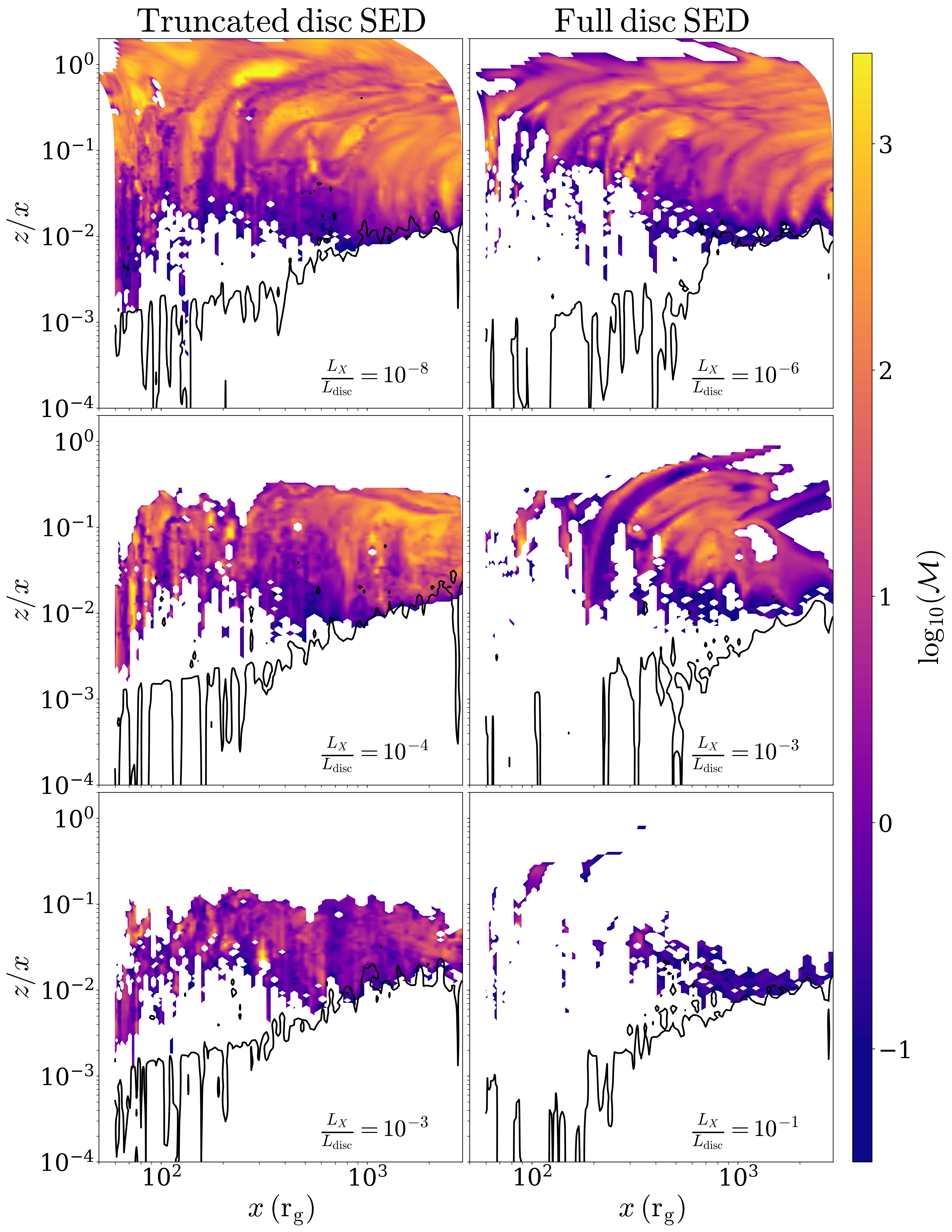}
    \caption{Force multiplier maps of 6 selected simulations as a function of $z/x$ and $x$. The left column shows the simulations with the truncated disc SED while the right column shows the simulations with the full disc SED. From top to bottom the X-ray luminosity of the central source increases. The black solid line gives the sonic point at each radius.}
    \label{fig:force_multiplier}
\end{figure}

We show on \autoref{fig:density_velocity} and \autoref{fig:force_multiplier} maps of the density, radial velocity and force multiplier for a set of 6 selected simulations that illustrate the behaviour of line-driven winds for different ratios of X-ray to disc luminosity, $L_X/L_\mathrm{disc}$. We also provide videos of these six simulations as supplementary material. We emphasize that what we call low, intermediate or high X-ray luminosity is different for the two sets of simulation with different disc SED.
We start by highlighting three features that are shared between all simulations, before describing individually each X-ray regime.

First, we see that all winds have a lot of small-scale structure. They have over-densities that can be as much as ten times denser than the background density as well as filaments that originate from the disc and propagate into the wind. This filamentary structure was noted as early as \cite{proga1998} and might be related to waves propagating at the surface of the disc and locally enhancing the mass-loss rate. We stress that this small-scale structure is not related to microscopic clumps such as can be produced by the line-deshadowing instability \citep{owocki1988} but is due to macro-clumps that emerge naturally from the simulation.

Second, we observe a difference of behaviour between the very low-density funnels of the truncated and full disc SED case. In the truncated SED case, the matter in the funnel is falling down while in the full disc SED case it is driven upwards. In both cases, the force multiplier is zero in the funnel, which is highly ionized. Hence, the tenuous wind in the funnel of the full disc SED case is driven by the radiation flux of the inner disc located between 6 $r_g$ and 60 $r_g$ interacting with matter through electron scattering only.

Third, we see from \autoref{fig:force_multiplier} that the force multiplier at the base of the wind from the innermost parts of the disc is always negligible, even above the sonic point. This means that the base of the wind is entirely accelerated by radiation pressure due to electron scattering and only at higher altitudes does line acceleration dominate. Hence, electron scattering is responsible for setting the mass-loss rate (at the sonic point) and line-driving determines afterwards whether the wind will successfully escape the system. To better understand this point and the role of electron scattering, we ran a simulation with force multiplier set to zero for the full disc SED (see video in the supplementary material). We find that electron scattering creates a disc atmosphere with recurrent small failed outflows being launched and quickly falling onto the disc. The mass-loss rate of this weak wind/atmosphere is 7 orders of magnitude lower than the accretion rate at the outer boundary of our simulation. Hence, this shows that the enhancement of the radiative force due to the force multiplier is a necessary ingredient in our simulation. 

We now detail the structure for each regime of X-ray strength.

\subsubsection{Low X-ray luminosities}
For low X-ray levels (top panels of \autoref{fig:density_velocity} and \autoref{fig:force_multiplier}), the wind is dense with number densities, $n=\rho/\mu m_p$, where we assume $\mu=0.6$ for simplicity, between $10^6$ and $10^9\:\mathrm{cm^{-3}}$ in its outer regions. It is also very fast with velocities between $10\%$ and half of the speed of light in the dense part of the wind and nearing the speed of light in the low-density funnel near the pole. The full disc SED wind is systematically faster than the truncated one across the entire domain. We see that, in general, the fastest part of the wind does not contribute significantly to the mass-loss. As can be seen from the polar plot on \autoref{fig:density_velocity}, half of the mass-loss is contained within the region between $\theta=[56^\circ,90^\circ]$ and $\theta=[61^\circ,90^\circ]$ for the simulation with the truncated disc SED and full disc SED respectively. Hence, the winds in the truncated case are more extended in $\theta$ than the full disc case, but neither covers the entire solid angle. The dependence of the geometry of the wind on the anisotropy of the driving radiation field was noted as early as \cite{proga1998}. For the full disc SED, the radiation field is more radial, while for the truncated disc SED, it is more vertical, so that the wind will extend closer to the pole in the latter case. The fact that the fastest regions of the wind do not coincide with the densest part of the wind means that the loss of kinetic energy occurs over a larger solid angle than the mass-loss rate, with half of the kinetic power loss located between $\theta$=[40°,90°] and $\theta$=[54°,90°] for the simulation with the truncated disc SED and full disc SED respectively.  Finally, we note that, especially for the truncated case, within the densest part of the wind there can be small regions where the velocity is very low or even negative, highlighting the complex dynamical structure of the wind.

\subsubsection{Intermediate X-ray luminosities}\label{sec:intermediate}

For intermediate X-ray levels (middle panels of \autoref{fig:density_velocity} and \autoref{fig:force_multiplier}), the wind is more localised in space and is very transient, as we have seen in \autoref{sec:massloss}. We show on the middle panels of \autoref{fig:density_velocity} and \autoref{fig:force_multiplier} the density, radial velocity and force multiplier during the episodic ejection events seen at 280 years and 110 years for the truncated and full disc SEDs respectively. Outside of these episodic ejection events, the wind is weak and mostly acts as a disc atmosphere with small ejections happening regularly. However, during these episodic ejection events the wind configuration is quite complex. For the full disc SED, we clearly see, at $\theta\approx \pi/4$ a filament of high density whose velocity at the base is negative. This is a transient dense wind that originated from the inner disc and was accelerated for a few years until its base started to fall down again (see video in the supplementary material). We see that this dense wind carries the entirety of the mass-loss rate and kinetic luminosity. This dense wind, which had a large acceleration at the time of its launch, is now escaping ballistically, but its base is falling down. Its initial acceleration stopped because of overionization by high-energy radiation, as attested by the small force multiplier within the dense filamentary flow. Behind this dense falling wind is another faster, more tenuous flow that is shielded from high energy radiation, as can be seen from its high force multiplier and high velocity. The dense wind provides a temporary shield for this component to be efficiently accelerated, but even this shielded wind will eventually be ionized and dies out once the failed wind falls back on the disc.

\begin{figure}
    \centering
    \includegraphics[width=90mm]{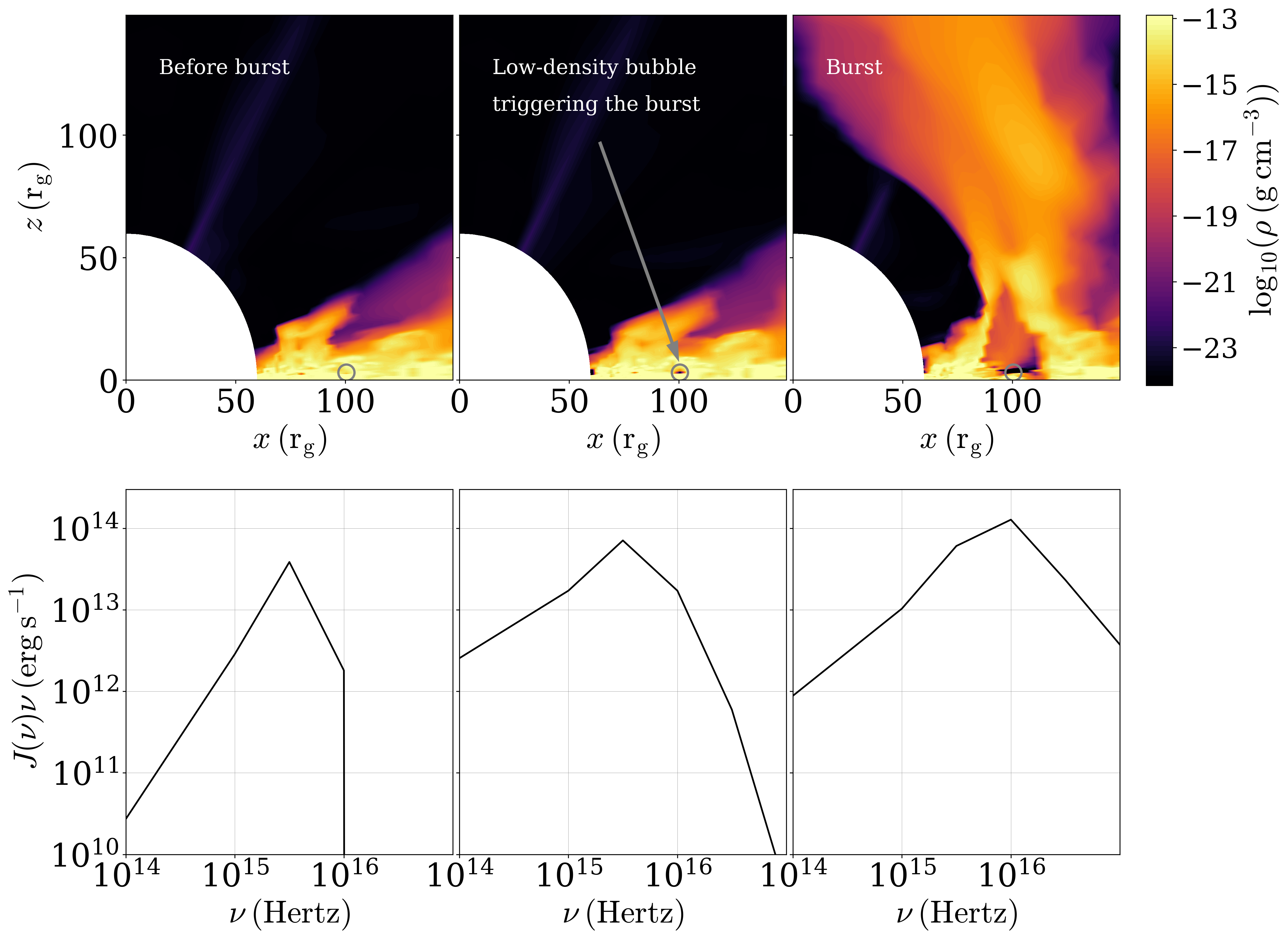}
    \caption{Top panels: Time series of density maps zoomed on the region where a transient ejection originates spanning the time before, during and after the transient ejection. Bottom panels: SEDs at the point circled on the top panels from which the transient ejection originated.}
    \label{fig:time_burst}
\end{figure}

To better understand how this specific shielding structure originates, we show on \autoref{fig:time_burst} a time series of snapshots spanning the time before, during and after the beginning  of the burst, for the simulation with a full disc SED and $L_X/L_\mathrm{disc}=10^{-3}$. The burst is triggered at the very base of the wind and starts with the formation of a low-density bubble within the disc atmosphere. When this bubble is created, both the spectral shape and the intensity of the driving flux passing through the cell where the bubble forms drastically change, increasing by a factor of 3 at its peak and having much more radiation at the high-frequency end of the UV band (see bottom panels of \autoref{fig:time_burst}). The changes in the driving flux SED allow for efficient acceleration, triggering the burst. We do not see any evidence that the creation of such low-density bubbles follows a well-defined cycle on long timescales. Instead, the bubbles appear to arise through stochastic fluctuations at the base of the wind, and although they are relatively common they often drive smaller ejections than that illustrated in \autoref{fig:time_burst}. Few seem to be able 1) to be shielded from intense radiation so as to have a high force multiplier and 2) to be large enough and close enough to the disc so as to cause large changes in the driving flux. 

\subsubsection{High X-ray luminosities}

For high X-ray levels (bottom panels of \autoref{fig:density_velocity} and \autoref{fig:force_multiplier}), the wind gets highly overionized by radiation. This can be seen from \autoref{fig:force_multiplier}, where we see that the regions where the force multiplier is non-zero get very restricted in space. Even where the force multiplier is non-zero, it is much smaller than at lower X-ray levels. As a result, for the truncated disc SED, we see small ejections forming but these are never able to escape, meaning that they form an extended atmosphere rather than a wind. For the full disc SED, large ejections occur for $L_X/L_\mathrm{disc}=10^{-1}$ at around 50 years and 150 years, but those ejections are rarer than at lower X-ray levels. Nonetheless, these ejections are quite massive and the transient shielding behaviour that we described for intermediate X-rays is also valid here. Finally, we also note that in the high and intermediate X-ray cases, an inflowing region forms near the disc at large radii -- a failed wind coinciding with regions of zero force multiplier.

\subsection{Force multiplier and ionic contributions}\label{sec:ion_contrib}

As we have seen in \autoref{sec:structure},  winds produced in these simulations can be massive and fast, either in a persistent way (as found at low X-ray levels) or in a transient way (as found at intermediate and high X-ray levels). The acceleration of such strong winds is enabled by the large force multipliers that are reached throughout the domain. Indeed, at low X-ray levels, a very large fraction of the domain has $\mathcal{M}>100$. However, as we increase the X-ray level we see that the force multiplier reached throughout the wind drastically decreases.

\begin{figure}
    \centering
    \includegraphics[width=90mm]{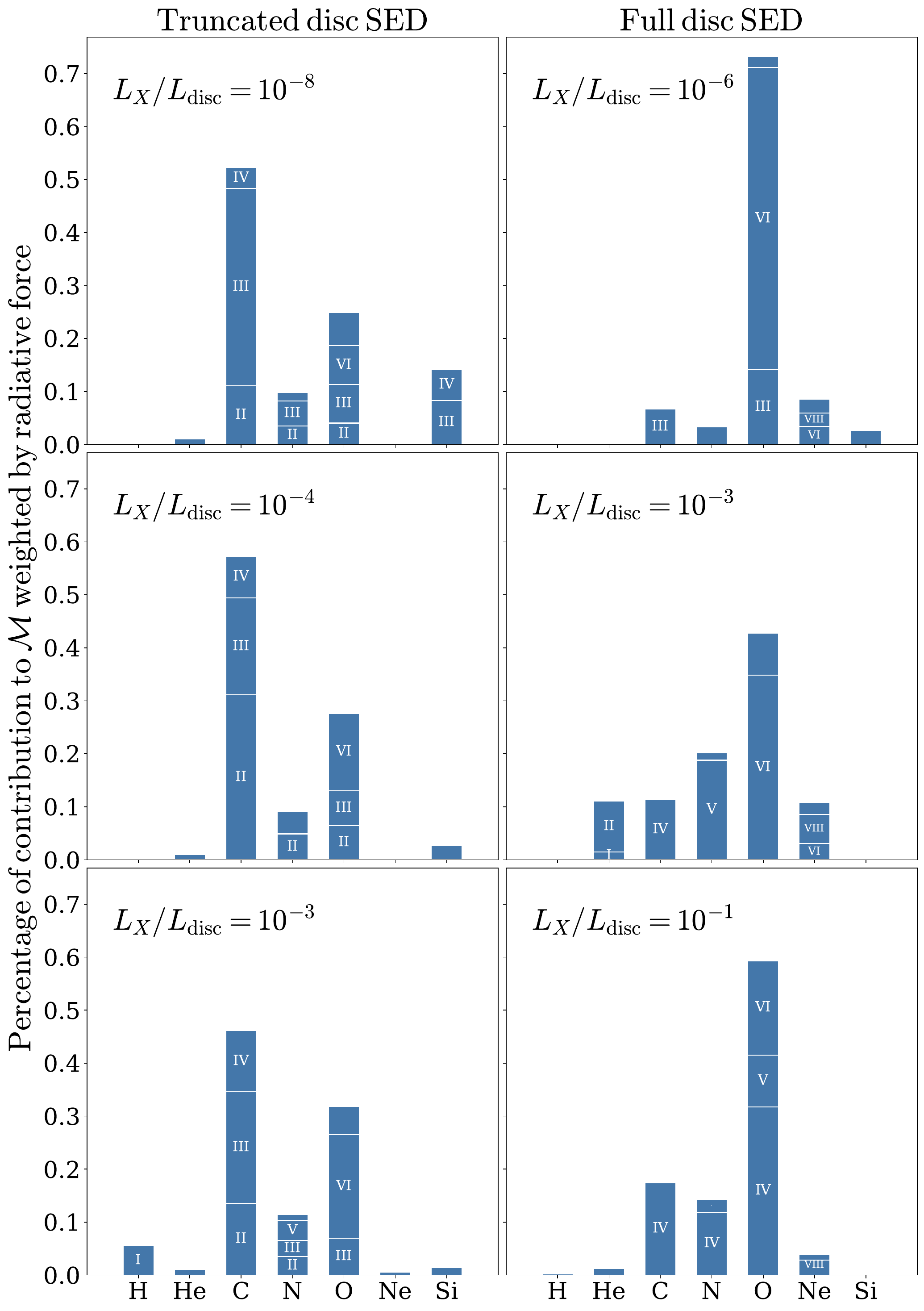}
    \caption{Percentage of contribution of ionic species to the acceleration of the wind. The left column shows the simulations with the truncated disc SED while the right column shows the simulations with the full disc SED. From top to bottom the X-ray luminosity of the central source increases.}
    \label{fig:ion_contrib}
\end{figure}

To better understand the force multiplier behaviour, we show in \autoref{fig:ion_contrib} the contribution of individual atomic species (see \autoref{eq:method_force_multiplier}) and their ionization stage to the force multiplier (see \autoref{eq:method_force_multiplier}) weighted by, $|\vec{g}_{\rm rad}|$, the magnitude of the radiative acceleration. Hence, the length of a bar on the histogram quantifies the contribution of an ionic species to the acceleration of the wind. We note that although electron scattering can account for significant acceleration at the base of the wind, it is not represented on the histogram. 

For the low X-ray case with the truncated SED, acceleration is dominated by many different atomic transitions associated to ionization energies below 50 eV such as \ion{C}{ii}, \ion{C}{iii}, \ion{N}{ii}, \ion{N}{iii}, \ion{O}{ii}, \ion{O}{iii}, \ion{Si}{iii} and \ion{Si}{iv} with far smaller contributions from higher energy lines from \ion{C}{iv} or \ion{O}{vi}. This is in contrast with the simulations using the full disc SED, where wind acceleration is mostly due to high ionization state ions, such as \ion{C}{iv}, \ion{O}{vi}, \ion{Ne}{vi} or \ion{Ne}{viii}. These differences are consistent with the fact that the input truncated disc SED only has significant radiation up to $20\:\mathrm{eV}$ (although reprocessing tends to redistribute radiation to slightly higher energies) while the full disc SED still has significant radiation up to $\approx$ 100 eV.

\begin{figure}
    \centering
    \includegraphics[width=90mm]{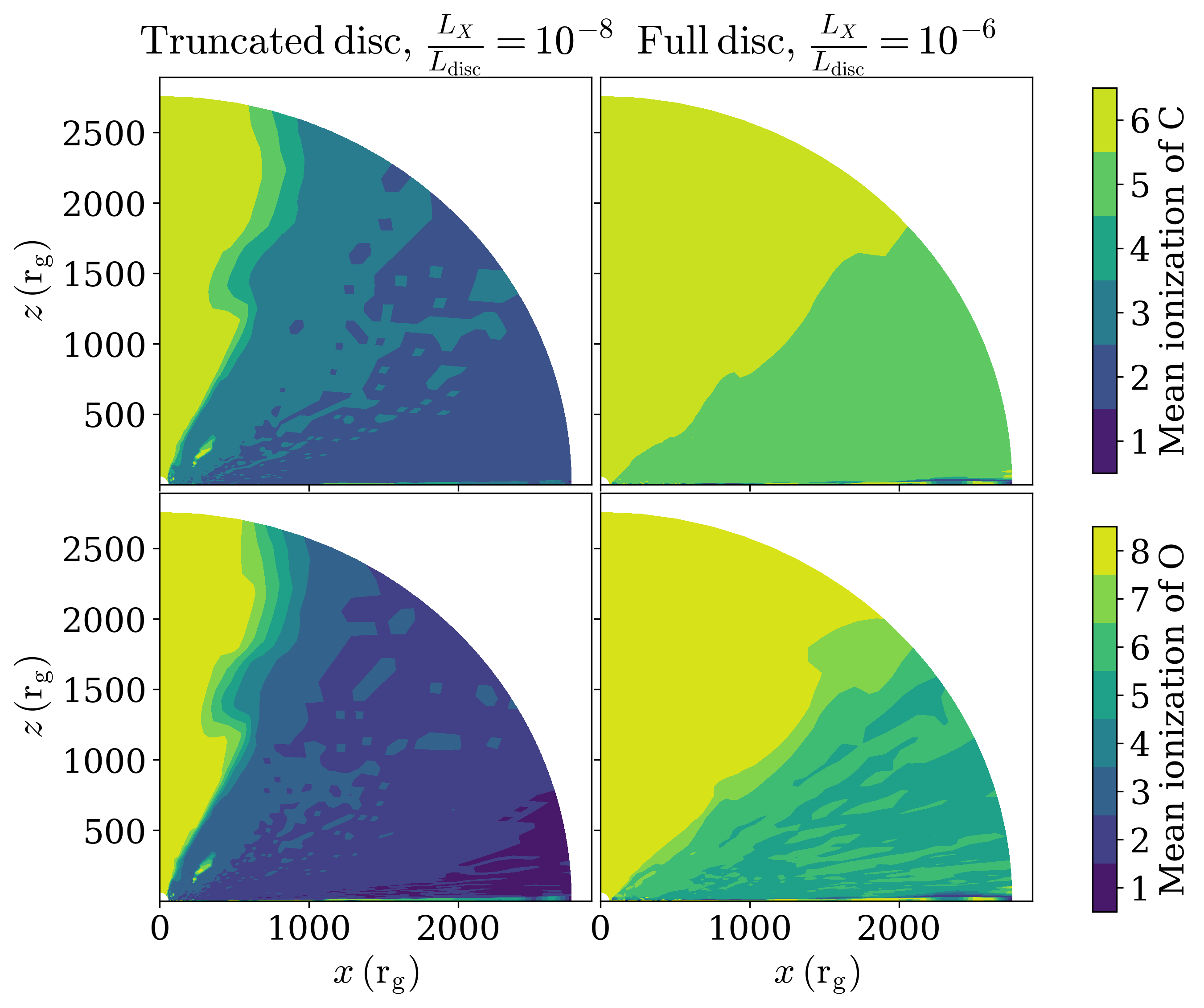}
    \caption{Mean ionization level of C (top panels) and O (bottom panels) for two simulations at low X-rays.}
    \label{fig:ion_abundances}
\end{figure}

However, it is perhaps surprising that the transitions with ionization energy around 50 eV or lower that are seen in the truncated disc SED case do not participate as well in the full disc SED case. To understand this, we have show in \autoref{fig:ion_abundances} maps of the mean ionization level of Carbon and Oxygen for the two low X-ray simulations with the two different disc SEDs. We see that the simulation with the full disc SED is much more ionized. Indeed, the mean level of Carbon in the truncated disc is \ion{C}{II} in the lower regions and \ion{C}{III} in the upper regions, while it is \ion{C}{V} and \ion{C}{VI} respectively for the full disc SED. The same trend exists for  Oxygen, which is dominated by \ion{O}{II} and \ion{O}{III} for the truncated case while it is dominated by \ion{O}{V} and \ion{O}{VI} for the full disc case. This means that the low energy ions disappear when using the full disc SED, explaining why their transitions do not contribute to the force multiplier. Whether these low energy transitions disappear because of overionization by the X-ray central source or the full disc SED is not exactly clear. We suspect that in the upper regions of the wind the X-rays dominate the overionization. However, in the densest part of the wind closer to the disc, we find that $\xi\equiv 4\pi F_X/n_H\approx 1$ so that the wind is not overly ionized by X-rays. Consequently, we suspect that the EUV disc emission is responsible for overionization near the disc. 

Surprisingly, we do not see a dramatic change in the contributing ionic species as we increase the X-ray luminosity (for the same disc SED). The force multiplier does decrease with increasing X-ray luminosity, because larger and larger parts of the domain are being overionized; however, for the regions of the wind that are efficiently driven, it is the same ionic species that participate in acceleration regardless of the X-ray luminosity. 

\subsection{Observational signatures}\label{sec:obs}

In this sub-section, to designate our simulations we replace our input parameter $L_X/L_\mathrm{disc}$ by the computed quantity $\alpha_\mathrm{OX}$ (see \autoref{eq:alphaOX}), where the luminosities are averaged over all solid angles. The correspondence between the input $L_X/L_\mathrm{disc}$ and emergent $\alpha_\mathrm{OX}$ is given in \autoref{tab:table1}.

\begin{table*}
    \centering
    \caption{Summary of the observational signatures of a selected set of simulations where we define $\alpha_\mathrm{_{OX}}$ as in \autoref{eq:alphaOX} with the luminosities being averaged over all solid angles. The X-ray levels (low, intermediate and high) are used to describe the wind dynamics in \autoref{sec:structure}, and match to top, middle and bottom columns, respectively, in Figures \ref{fig:density_velocity},~\ref{fig:force_multiplier},~\ref{fig:ion_contrib},~\ref{fig:spectra} and \ref{fig:spectra_angle}.
    }
    \setlength\tabcolsep{0pt}
    \begin{tabular*}{\linewidth}{@{\extracolsep{\fill}} cccccc }
        \midrule
        \midrule
        $L_X/L_\mathrm{disc}$ (input) & $\alpha_\mathrm{_{OX}}$ (emerging) &  X-ray level  & \textbf{FeLoBAL} &  \textbf{HiBAL} & \textbf{UFO} \\

        \midrule
        \midrule
            \multicolumn{5}{c}{\bfseries{Full disc SED}} \\
        \midrule
         $10^{-6}$ & -5.98 & ``Low'' & No & for $i\in[70°;85°]$ & -  \\  
         $10^{-3}$ & -2.88 & ``Intermediate'' &No & for $i\in[65°;85°]$ & for $i\in[50°;60°]$ \\
         $10^{-1}$ & -0.74 & ``High'' &No & No & for $i\in[55°;70°]$\\
        \midrule
         \multicolumn{5}{c}{\bfseries{Truncated disc SED}} \\
        \midrule
         $10^{-8}$ & -8.08 & ``Low'' & for $i\in[55°;85°]$ & No & - \\  
          $10^{-4}$ & -3.97 & ``Intermediate'' & for $i\in[80°;85°]$ & No & - \\ 
          $10^{-3}$ & -2.92 & ``High'' & No & No & - \\ 
        \midrule
    \end{tabular*}
    \label{tab:table2}
\end{table*}

\begin{figure*}
    \centering
    \includegraphics[width=\textwidth]{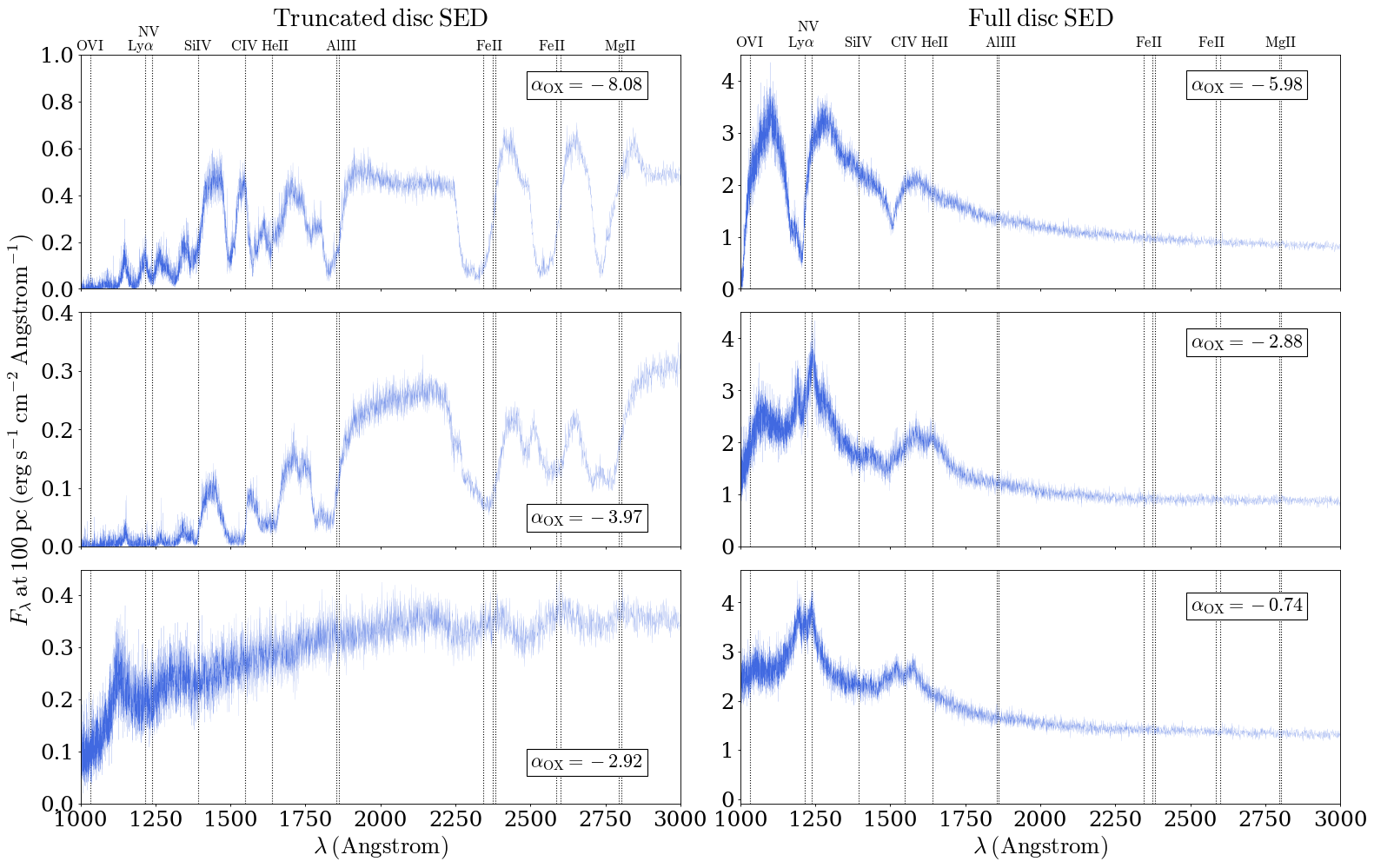}
    \caption{UV spectra at an inclination of $80^\circ$ with all relevant line transitions overlaid. The left column shows the simulations with the truncated disc SED while the right column shows the simulations with the full disc SED. From top to bottom the X-ray luminosity of the central source increases.}
    \label{fig:spectra}
\end{figure*}

\begin{figure*}
    \centering
    \includegraphics[width=\textwidth]{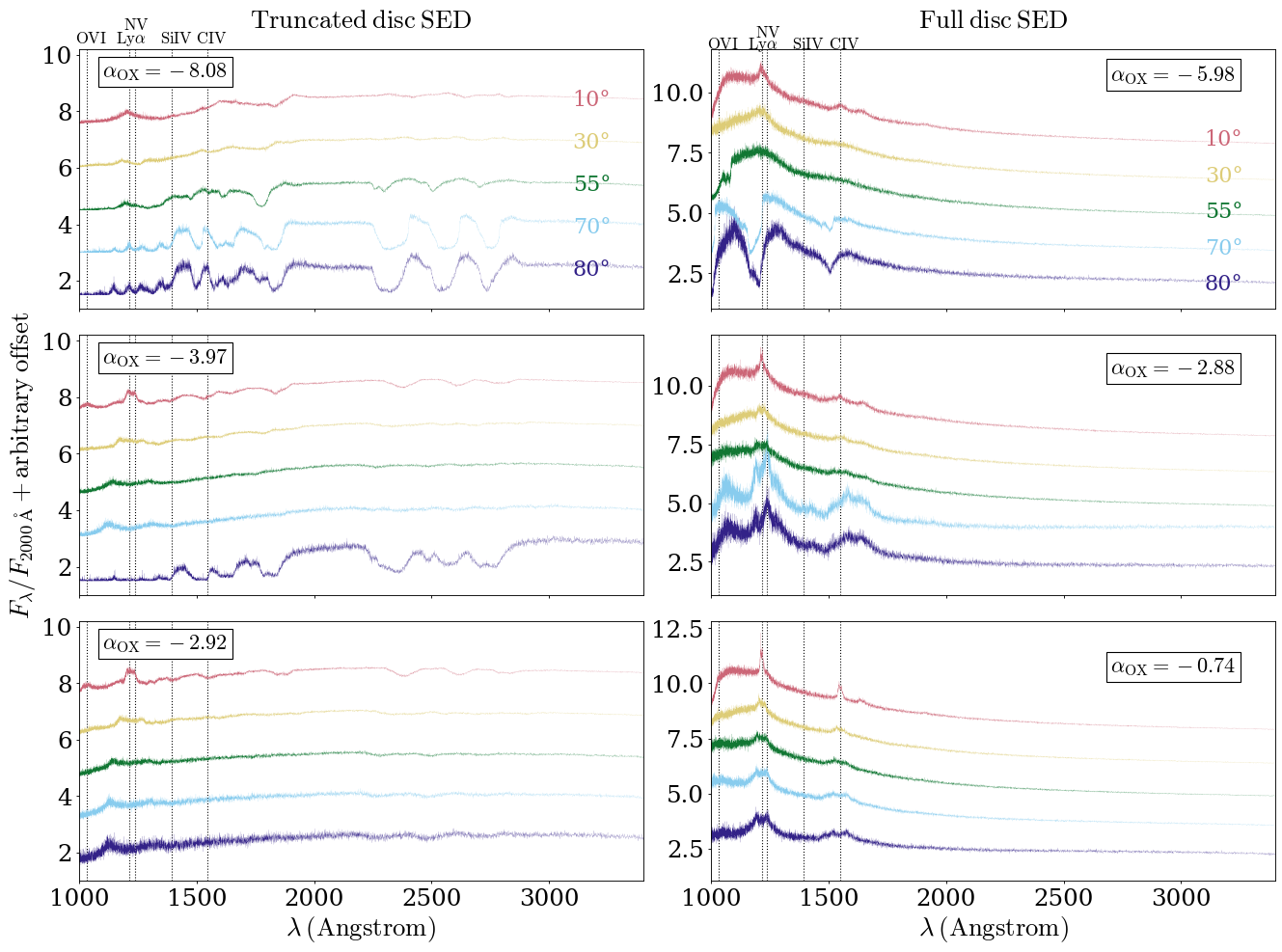}
    \caption{UV spectra at inclinations of $80^\circ$, $70^\circ$, $55^\circ$, $30^\circ$ and $10^\circ$. The left column shows the simulations with the truncated disc SED while the right column shows the simulations with the full disc SED. From top to bottom the X-ray luminosity of the central source increases.}
    \label{fig:spectra_angle}
\end{figure*}

In \autoref{fig:spectra}, we present UV  spectra from our six simulations at an equatorial inclination angle of $80^\circ$. We also show in \autoref{fig:spectra_angle}, the spectra for $80^\circ$, $70^\circ$, $55^\circ$, $30^\circ$ and $10^\circ$ for our six selected simulations. Finally, we summarize in \autoref{tab:table2} the observational properties of our simulations. We note that to compute the spectra, we relax the isothermal assumption made throughout this paper and let SIROCCO self-consistently compute the temperature during the ionization calculations (see \autoref{appendix:isothermal}). This avoids artefacts in the spectra due to the isothermal constraint. 

The most striking result is that, for our weakest X-ray runs ($\alpha_\mathrm{_{OX}}=-5.98$ and -8.08 for the full and truncated disc SED respectively), we are able to produce spectra resembling HiBAL and FeLoBAL quasars. Although the X-ray level is unrealistically weak compared to observed quasars, this shows that once our winds are powerful enough, they can produce some observed features of BAL quasars. Interestingly, we see that the HiBAL and FeLoBAL features are also visible in the transient winds seen at intermediate X-ray levels ($\alpha_\mathrm{_{OX}}=-2.88$ and -3.97 for the full and truncated disc SED respectively), although they tend to be weaker (especially for the full disc SED). However for the highest X-ray runs ($\alpha_\mathrm{_{OX}}=-0.74$ and -2.92 for the full and truncated disc SED respectively) the BAL features seem to have disappeared completely although we do still see some weak absorption features in \ion{Fe}{II} and \ion{Mg}{II} and possibly \ion{O}{VI} for the truncated disc SED.

FeLoBAL features -- deep absorption troughs due to \ion{C}{IV}, \ion{Al}{III}, \ion{Fe}{II} and \ion{Mg}{II} -- are only seen for the truncated disc SED, at low X-ray luminosity and at large inclination angles larger than 55 degrees. These features originate from the densest, slowest part of the wind at large radii. Interestingly, we see that although \ion{Mg}{II}, \ion{Al}{III} or \ion{Fe}{II} atomic lines dominate the spectrum they do not contribute significantly to the acceleration of the wind as they are weak lines. 

Conversely, HiBAL features -- strong absorption features due to \ion{C}{IV} and \ion{N}{V} (and also a faint signature of \ion{Si}{IV}) -- are only seen for the full disc SED at high inclination angle above 70 degrees and for $\alpha_\mathrm{_{OX}}=-5.98$ and $\alpha_\mathrm{_{OX}}=-2.88$ (although they are less prominent for the latter). These features originate from the fastest regions of the wind. It is interesting to note that in the case of the failed transient winds seen for $\alpha_\mathrm{_{OX}}=-2.88$, the dense shielding wind does not participate to the BAL signature as it is too ionized. Only the regions that are shielded from X-rays produce a HiBAL signature. Furthermore, for the full disc SED with $\alpha_\mathrm{_{OX}}=-0.74$ (bottom right panel), we do not see any BAL features although the transient ejection is quite massive. In fact, we computed the spectrum from both ejection events that can be seen on \autoref{fig:massloss} and none of them is associated with a BAL signature. This is most likely because the wind is too ionized, even behind the failed shielding wind. Hence, launching successfully a wind in a highly X-ray rich environment does not necessarily imply observed BAL features. However, as we will see in \S\ref{sec:UFO}, the runs with the strongest X-rays can produce absorption features in lines from species with higher ionization energies.

Finally, we note that as we decrease the inclination angle, all simulations tend to go from deep absorption troughs to shallower ones and eventually to emission lines. Moreover for the highest X-ray emission runs, emission lines (mostly \ion{Ly}{$\alpha$} and \ion{C}{IV}) are seen at all angles going from double-peaked to single-peaked as we decrease the inclination angle. 

We discuss the consequences of our results for current BAL unification models in \S\ref{sec:BAL}. We also compiled these results in \autoref{tab:table2} along with the results from \S\ref{sec:UFO}.

\section{Discussion}\label{section:discussion}

\subsection{Comparison to previous simulations}\label{sec:comparison}
Our simulations differ from most of the literature on AGN line-driven winds in two aspects. First, our use of Monte Carlo radiative transfer means we consider a more realistic treatment of radiation that is multi-frequency and multi-directional, allowing to us take into account UV and X-ray attenuation, scattering and reprocessing; by contrast, most of the literature consider that the UV disc emission propagates through an optically thin wind and that the X-rays are attenuated but not scattered or reprocessed \citep{proga2000,proga2004,nomura2016,mizumoto2021}. Second, we compute the force multiplier from an iterative Monte-Carlo calculation of the ionization structure of the wind taking into account the full local SED of radiation, instead of relying on a force multiplier prescription as a function of the ionization parameter $\xi$, which depends only on the X-ray flux in a given cell. 

In previous simulations using simplified radiative transfer and a force multiplier prescription, it was found that winds can self-shield from the X-ray radiation and produce steady winds that are quite powerful with mass loss rates $\approx 15\%$ the accretion rate for a $10^8\:M_\odot$ black hole accreting at 0.5 Eddington \citep{proga2000,proga2004,nomura2016} and as high as $\approx 2$ times the accretion rate for a $10^9\:M_\odot$ black hole accreting at 0.5 Eddington \citep{nomura2017}. This self-shielded structure, with a slow and dense inner wind shielding the outer fast wind is now well-established in this type of simulations. However, these simulations provide an ideal case where the X-ray flux is maximally attenuated and the UV flux is maximized. 

A notable improvement was made recently by \cite{dyda2024} where the authors performed radiative hydrodynamics simulations taking into account absorption, scattering and re-processing of the X-rays. As before, the authors report that with pure attenuation of the X-rays the wind forms the self-shielded structure. However, they ran their simulations for longer than \cite{proga2004} and found that their wind is intermittent with a quasi-period of $\approx 100$ inner orbital periods. This quasi-period is close to ours, which is $\approx 200$ inner orbital periods for our simulation with a full disc SED and $L_X/L_\mathrm{disc}=10^{-3}$. The authors attribute this behaviour to a cyclic building/ejection of mass in the inner parts of the disc. When enough mass has built in the inner disc, shielding can occur for a wind to be launched. The period between wind ejection is then the time it takes for the wind to die out and for mass to build up again in the disc. We tried to interpret the behaviour of our wind within this scenario. However, we do not see a strong correlation between mass build up in the inner parts at the base of the wind and efficient wind acceleration. As we discuss in \autoref{sec:structure}, it seems that the ejections are due to a much more localized, stochastic event than a global, coherent process. In any case, the crucial result of \cite{dyda2024} is that including scattering and reprocessing of the X-rays weaken the wind and can even suppress it entirely, in agreement with our work. This result again stresses the importance of proper radiative transfer calculations.

\begin{figure}
    \centering
    \includegraphics[width=90mm]{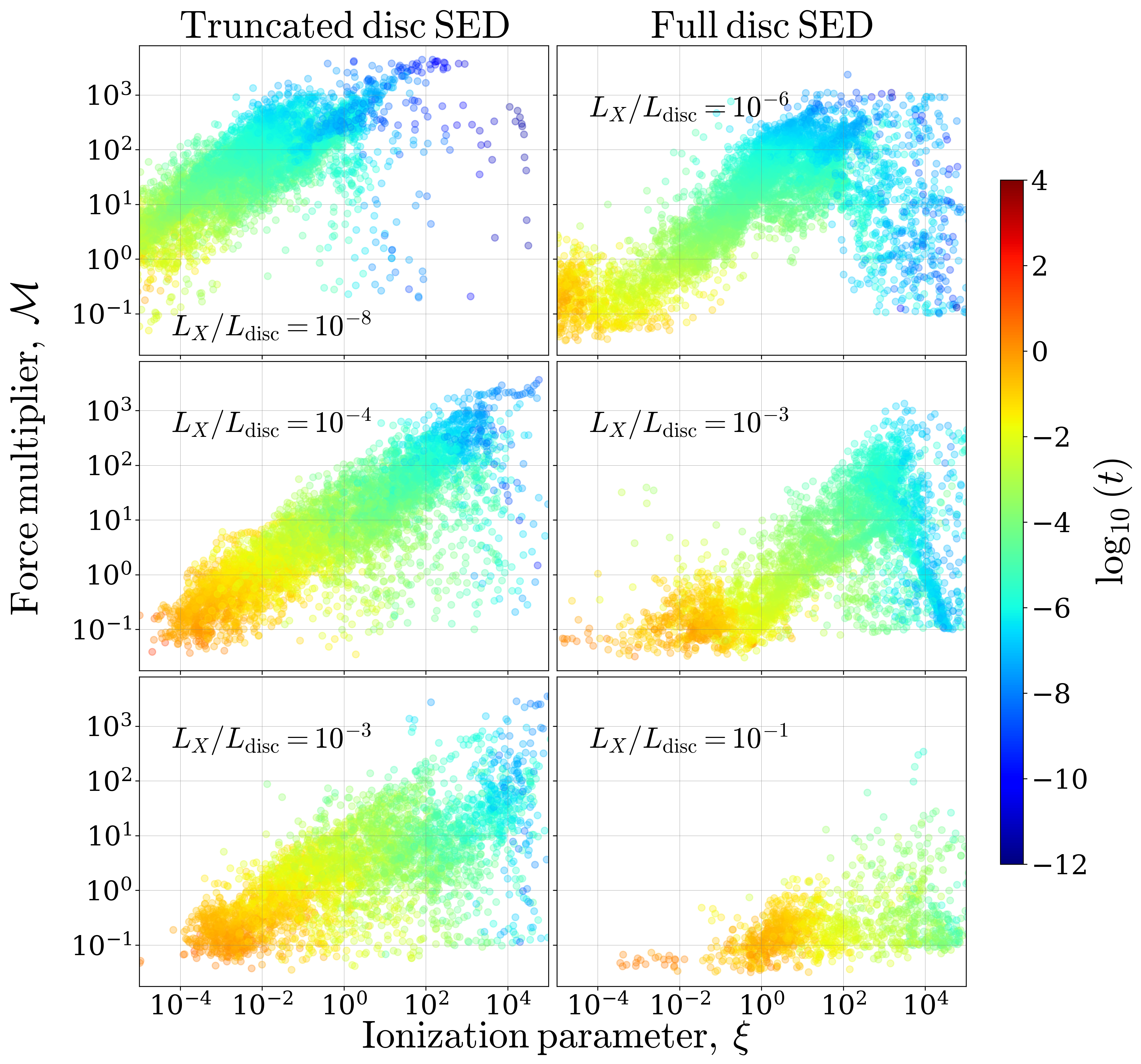}
    \caption{Scatter plot of the force multiplier as a function of the ionization degree colored by the Sobolev optical depth. The left column shows the simulations with the truncated disc SED while the right column shows the simulations with the full disc SED. From top to bottom the X-ray luminosity of the central source increases.}
    \label{fig:M_xi}
\end{figure}

\cite{dyda2025} also recently investigated the effect of using a more sophisticated treatment of the force multiplier by using the force multiplier tables from \cite{dannen2019}. Notably, \cite{dannen2019} found that X-ray lines can participate in line-driving at a level comparable to UV lines. This allows a force multiplier of 10-100 to be present up to ionization parameters as high as $\xi \approx 10^3$. With this new treatment of the force multiplier, \cite{dyda2025} reported that their line-driven wind is more powerful than before because of this extra driving due to X-ray lines, even for X-ray levels as low as $L_X/L_\mathrm{disc}=0.05-0.1$. In the simulations of our present paper, we do not allow for X-ray driving, which will be studied in future papers. Nonetheless, we find that we can also have a relatively high force multiplier (${\cal M}\gtrsim100$) even for $\xi\approx 10^3$. We show in \autoref{fig:M_xi} scatter plots of $\mathcal{M}$ as a function of $\xi$ with the color showing the Sobolev optical depth $t$. For a dimensionless optical depth of $t\approx 10^{-8}$ we often have $10^2<\mathcal{M}<10^3$ even for $\xi\approx 10^3$. For a given simulation, $\mathcal{M}$ roughly follows $\xi$, albeit with a large dispersion, suggesting that we could in principle find a unique formula for $\mathcal{M}(t,\xi)$. However, we see that the form of $\mathcal{M}(t,\xi)$ is different for each simulation, strongly suggesting that $\mathcal{M}$ is not a function of $\xi$ and $t$ only. As a result, computations of the force multiplier from the full ionizing flux SED, or at least more data than simply the X-ray flux, are crucial.

\subsection{Observational implications}

\subsubsection{Broad-absorption line quasars}\label{sec:BAL}

There are two existing paradigms for explaining the occurrence of BAL quasars within the quasar population. One proposes that BAL and non-BAL quasars represent different evolutionary stages of an AGN \citep{voit1993}. The other proposes that, on the contrary, BAL and non-BAL quasars belong to the same underlying population but are seen from different inclinations  \citep{murray1995,elvis2000,matthews2016}. In the latter case, AGN that are seen almost edge-on have the strongest absorption signatures, including low ionization lines that are buried deep in the wind, which would correspond to FeLoBALs. AGN that are seen at more moderate inclination would exhibit absorption in the strongest resonance lines from regions that are more ionized, corresponding to HiBAL quasars. Finally, AGNs seen at low inclination angles would exhibit strong emission lines, corresponding to a BEL quasar. These `orientation' and `evolution' scenarios are not mutually exclusive nor a clear dichotomy, since the spectra are likely to both evolve with time and change with viewing angle \citep{giustini2019}. 

\autoref{fig:spectra_angle} and \autoref{tab:table2} suggest that the geometric/orientation BAL unification scheme does hold to a certain extent. Our simulations showing a HiBAL quasar signature at high inclination angles ($70^\circ$ or $80^\circ$) exhibit more of a BEL signature at low inclination angles ($<55^\circ$). However, we find that simulations looking alike a FeLoBAL quasar at high inclination angle still show absorption signatures (although much weaker) at lower inclination angle. Moreover, we find that the cases showing the clearest BEL signatures are simulations with high $\alpha_\mathrm{_{OX}}$, regardless of inclination, with emission lines going from single-peaked to double-peaked as the inclination angle increases. 

Unlike the models developed by \cite{elvis2000}, we find that in our simulations the formation of a HiBAL quasar or FeLoBAL quasar is not only dictated by the inclination angle. It is the SED of the disc that is the critical criterion for whether the wind produces a FeLoBAL or a HiBAL quasar signature. Indeed, we see FeLoBALs when the disc SED peaks at $8\times 10^{14}$ Hz while we see HiBALs when the disc SED peaks at $4\times 10^{15}$ Hz. This is particularly interesting in the context of other works that suggest an anti-correlation between the hardness of the EUV spectrum (as measured through the \ion{He}{ii}~1640\AA\ equivalent width) and the presence of BAL features or a strongly blueshifted \ion{C}{iv} 1550\AA\ emission line \citep{richards2011,rankine2020,temple2023}. While this trend is often interpreted as the fact that winds get weaker for harder EUV \citep{richards2011}, we see that there can be another interpretation. We find that the lowest X-ray runs with the two different SEDs have mass-loss rates that are very similar yet their signature is very different. Hence, it is not only the strength of the wind (its mass-loss rate and power) but also the SED illuminating the wind that determines its ionization structure and the observational signatures it ultimately produces. 

We also find that as we increase the strength of the X-rays the BAL signatures (whether of a HiBAL or a LoBAL signature) get weaker and more transient until they disappear for our highest X-ray strength. This is consistent with observations showing that the population of BAL QSOs tends to have a lower ratio of X-rays to UV luminosity, possibly intrinsic to the source (\citealt{green1995,gallagher2006,giustini2019} although see \citealt{hiremath2025}). 

Finally, our simulation with $L_X/L_\mathrm{disc}=10^{-3}$ and a full disc SED shows a transient signature of a HiBAL during the flare in mass-loss rate. The recurrence time of this signature in our simulation is roughly 100 days and the change between signatures happen over a timescale of years. There are quasars that have been observed to switch between BAL and non-BAL states on timescale of years, which would be consistent with our simulation \citep{filizak2012,mcgraw2017}. That said, if all detected BALs originated from a transient wind, such a large recurrence time scale would imply a much lower rate of BAL detection than observed. We believe that the recurrence time scale of wind ejection events in our simulations is tightly linked to the treatment of the disc-wind connection, and therefore delay any further comparison of inferred detection rates of BALs until this treatment is developed further.

\subsubsection{Ultra-Fast Outflows}\label{sec:UFO}

Ultra-Fast Outflows (UFOs) are detected through blue-shifted absorption lines from iron K shell transitions in the X-ray band between 2 and 10 keV. UFOs have large velocities that range between $0.03$ and $0.3\:c$ so that they are expected to be launched from the innermost regions of the disc \citep{tombesi2012}. They are typically thought to originate from a faster wind than the one of BAL quasars (although see \cite{vietri2022} for an ultra-fast BAL quasar). UFOs have been seen in many type of local AGN, going from radio-loud to radio-quiet quasars \citep{tombesi2010,tombesi2014}. However, simultaneous detection of BAL and UFO signatures have not been reported so far so that it is not clear if UFO and BAL are representing two different AGN populations.

\begin{figure}
    \centering
    \includegraphics[width=90mm]{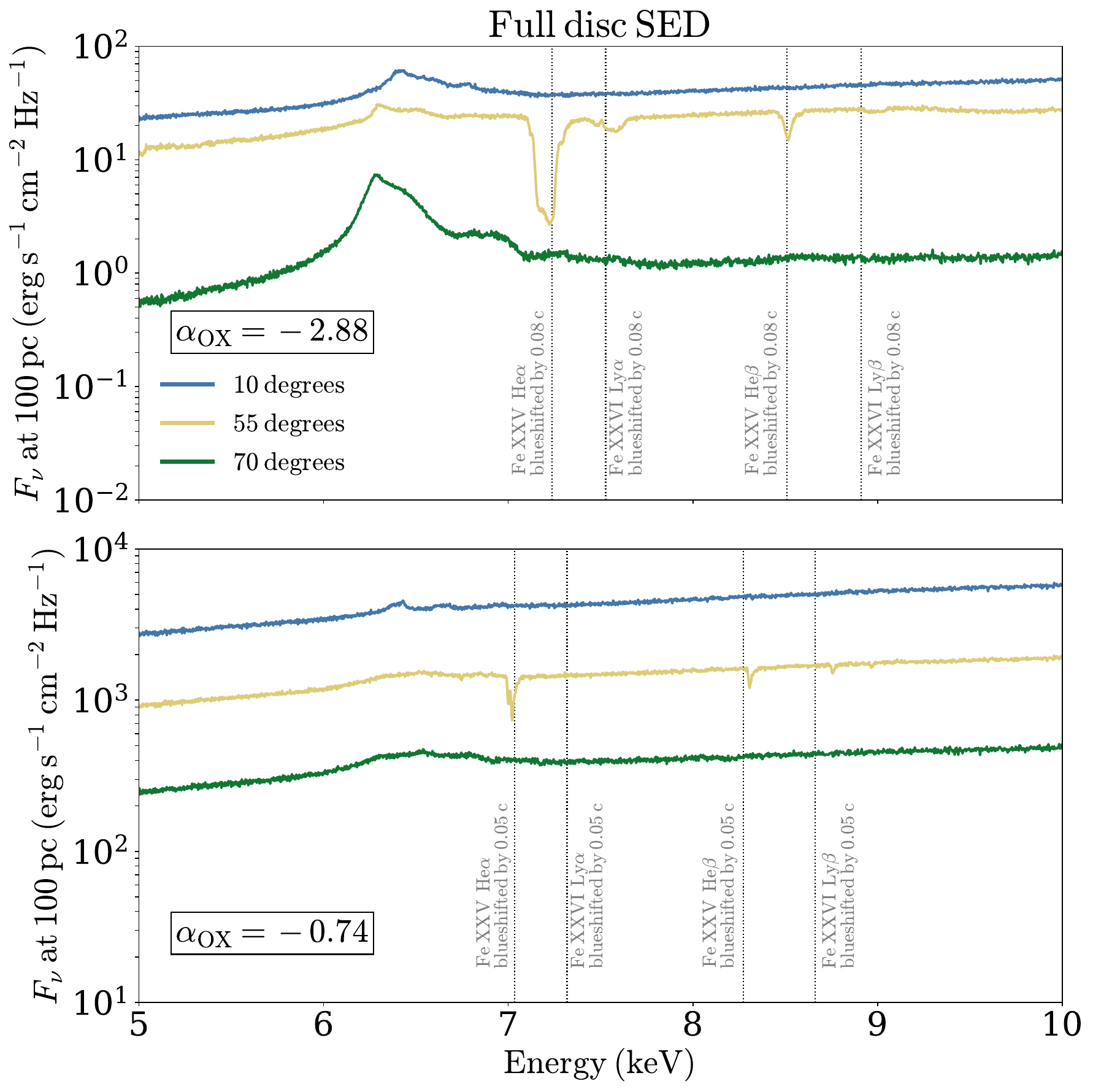}
    \caption{X-ray spectrum between 6 and 10 keV for our full disc simulations with $L_X/L_\mathrm{disc}=10^{-2}$ and $10^{-1}$ for 3 different inclination angles. We also plotted the \ion{Fe}{XXV} \ion{He}{$\alpha$}, \ion{Fe}{XXVI} \ion{Ly}{$\alpha$}, \ion{Fe}{XXV} \ion{He}{$\beta$} and \ion{Fe}{XXVI} \ion{Ly}{$\beta$} lines blue-shifted by 0.08 and 0.05 $c$ for $\alpha_\mathrm{_{OX}}=-2.88$ and -0.74 respectively.}
    \label{fig:spectra_Xray}
\end{figure}

We plot on \autoref{fig:spectra_Xray} the X-ray spectrum\footnote{Note that we used the fe$\_$17to27.dat atomic data set from SIROCCO to compute the X-ray spectrum.} for three inclination angles for the simulation with a full disc SED, $\alpha_\mathrm{_{OX}}=-2.88$ and -0.74. We see that we can produce absorption quite narrow absorption lines that could correspond to \ion{Fe}{XXV} \ion{He}{$\alpha$}, \ion{Fe}{XXVI} \ion{Ly}{$\alpha$}, \ion{Fe}{XXV} \ion{He}{$\beta$} and \ion{Fe}{XXVI} \ion{Ly}{$\beta$} blue-shifted by roughly 0.08 and 0.05 $c$ for $\alpha_\mathrm{_{OX}}=-2.88$ and -0.74 respectively. We applied a uniform blue-shift for all lines but we see that the absorption features already reveal a more complex wind structure with components having slightly different velocities. This could be of interest in the context of recent observations by XRISM that revealed that the broad blue-shifted absorption lines that were observed with previous X-ray instruments were actually composed of several (five or six) sets of narrow blue-shifted absorption lines having different velocities \citep{xrism2025,mizumoto2026}, which were interpreted as multiple clumps of dense material along our line of sight. Our simulation is quite in line with this picture, as it produces a set of narrow absorption lines originating from a dense discrete filamentary structure. However, where observations reveal five or six set of absorption lines with velocities ranging from $0.07$ $c$ to $0.4$ $c$ \citep{xrism2025,mizumoto2026}, our spectrum show only one set of absorption lines at roughly 0.08 and 0.05 $c$ for $\alpha_\mathrm{_{OX}}=-2.88$ and -0.74, respectively, reflecting the presence of only one dense filament at a time in our simulations. As we discuss in \ref{sec:intermediate}, the filaments in our models originate from close to the disc surface, which we treat as a boundary condition. A more realistic set-up with a turbulent disc-wind interface, as observed in \cite{jacquemin2020}, might give rise to the creation of multiple filaments with different velocities, so that the number of filaments (and set of absorption lines) in our simulations should be taken with a grain of salt.

As for the BAL signature for $\alpha_\mathrm{_{OX}}=-2.88$, the UFO signature is very localized in angle as we only see it for a line-of-sight of $55^\circ$. This is because it originates from the dense, over-ionized failed wind seen on \autoref{fig:density_velocity}, which is itself very localized in space. It is interesting to note that the BAL signature in this simulation was not coming from this failed, shielding wind which was too ionized but rather from the faster but less dense shielded wind below it. Conversely, the fastest shielded wind is not ionized enough to provide a UFO signature. Hence, our wind naturally produce a very clear distinction between BAL signatures and UFO signatures depending on the line-of-sight. For our full disc SED with $\alpha_\mathrm{_{OX}}=-0.74$, we also find that the transient wind is able to produce a UFO signature although it is too ionized to produce a BAL signature. This shows that the level of X-ray irradiation is also a critical discriminant between what would be observed as a UFO or a BAL AGN. Finally, we also note that our wind is able to produce broad iron emission lines for all inclination angles. This could be important for model of iron line reflection in the context of spin measurements. We plan to explore in more depth the implication of our simulations on X-ray signatures of AGN winds in a future paper.

\subsubsection{Changing-Look AGN}
Changing-look AGN (CL AGN) are a type of AGN evolving on very short time scales (typically a few months to years) compared to the viscous time scales (typically million of years) \citep{lawrence2018}. CL AGN are historically linked to the appearance or disappearance of broad emission lines in \ion{H}{$\alpha$} and \ion{H}{$\beta$} coming from the so-called broad-line region (BLR) but many are now observed to also have a changing continuum in the UV, optical and sometimes even X-rays \citep{tohline1976,lamassa2015,macleod2016,trakhtenbrot2019,ricci2020}. It is now accepted that some if not the majority of CL AGN are driven by a change in the accretion rate rather than an obscuring effect by a cloud of gas through the line of sight \citep{runnoe2016,trakhtenbrot2019}. However, it is not clear if the appearance/disappearance of the broad emission lines is due to the creation/destruction of the BLR or simply to the fact that the BLR is revealed/hidden by an increase/decrease in the continuum UV flux. Since CL AGN surveys focus on optical wavelength they report mostly the \ion{H}{$\alpha$} and \ion{H}{$\beta$} line evolution. However, recent studies on high redshift quasars have discovered \ion{Ly}{$\alpha$} and \ion{C}{IV} CL AGN \citep{ross2020,guo2025} so that UV and optical lines might trace the same phenomenon.

In our simulations, we find that, at high X-ray luminosities, a transient ejection event can suddenly produce double-peaked emission lines in \ion{Ly}{$\alpha$} and \ion{C}{IV}. The timescale on which these emission lines are produced is of the order of one year, very reminiscent of the typical time scale involved in CL AGN. Hence, we propose here that transient ejection events could produce behaviour reminiscent of a CL AGN. We further extrapolate that such fast ejection events should affect the disc mass accretion rate. Indeed, we find that these ejection events have a mass loss rate of the order of the accretion rate and happen on a time scale much faster than the local viscous time scale. Hence, the disc will not have time to adjust to this loss of mass, creating a deficit of mass at a given radius, that might translate to a signature in the observed continuum. For the typical radii at which we see these ejection events forming (typically $\approx 100\:r_{\rm g}$) this would show up as a deficit of optical emission. Hence, we propose that CL AGN events produced by transient ejection should show up as a Type 2 to Type 1 transition with a decreasing optical flux and could potentially explain the apparent decorrelation between disc emission and BLR emission observed in \cite{dehghanian2019}. While our simulations only produces changes in \ion{Ly}{$\alpha$} and \ion{C}{IV} it is plausible that while propagating outwards and cooling this ejection event will eventually produce \ion{H}{$\alpha$} and \ion{H}{$\beta$} lines although this clearly requires justification.

\section{Conclusions}\label{section:conclusion}

We have performed a set of 10 radiation-ionization hydrodynamic simulations to study line-driven winds in a $10^9\:M_\odot$ AGN accreting at $\approx 0.66 \dot{M}_\mathrm{Edd}$ under the influence of a central source of X-rays. We used two different SEDs for the emission from the disc, one originating from a disc truncated at $60\:r_{\rm g}$ and one from a disc extending down to $6\:r_{\rm g}$. We also varied the strength of the X-ray central source going from $L_X/L_\mathrm{disc}$ as low as $10^{-8}$ and as high as $10^{-1}$. By coupling the hydrodynamical code PLUTO and the Monte-Carlo radiative transfer and photo-ionization code SIROCCO, we were able to perform the first AGN line-driven wind simulations that self-consistently compute the dynamics, the multi-frequency and multi-directional radiative transfer and the non-LTE photoionization state of the plasma. Our conclusions are as follow: 
\begin{itemize}
\item At low X-ray luminosities, line-driving can produce strong, steady winds with high mass loss rates of $\approx20\%$ of the accretion rate covering half of the domain in latitude. Interestingly, the full disc SED and truncated SED produce similar mass-loss rates (despite the former being ten times more luminous) because of over-ionization of the wind by the disc SED in the full disc case.

\item The strong steady winds produced in some our simulations can produce either FeLoBAL or HiBAL features at high inclination for the truncated and full disc SED respectively. Interestingly, FeLoBAL and HiBAL trace different disc SEDs rather than different inclination angles. At low inclination, strong steady winds produce broad emission lines in \ion{Ly}{$\alpha$} and \ion{C}{IV}.

\item At high X-ray luminosities approaching realistic values of $L_X/L_\mathrm{disc}>10^{-3}$ (or $\alpha_{\rm _{OX}}>-3$), the outcome depends on the disc SED. For the truncated disc SED, the wind cannot survive for $L_X/L_\mathrm{disc}\ge 10^{-3}$. For the full disc SED, the wind can survive up to $L_X/L_\mathrm{disc}=10^{-1}$ ($\alpha_{\rm _{OX}}=-1$), which is the maximum level of X-rays we have tried. However, at high X-ray luminosities the wind becomes transient alternating between long periods (of a few hundred of years) of weak mass loss and short period (of a few years) of intense mass loss with the mass-loss rate being of the order or exceeding the accretion rate and being very localized in solid angle.

\item The transient winds produced in simulations with $\alpha_\mathrm{_{OX}}<-2$ can also produce FeLoBAL and HiBAL features during their intense ejection phase. However, the BAL signature is very localized in space and fainter. We also find that the high X-ray runs can produce a UFO signature but at a different angle than the BAL signature. The transient wind simulation with $\alpha_\mathrm{_{OX}}=-1$ also produces a UFO signature although it does not produce a BAL signature because it is too ionized to do so. Hence, our simulation always show a dichotomy between BAL and UFO signatures whether with the inclination angle or with the X-ray level.

\end{itemize}

To conclude, we find that the over-ionizing problem in AGN is far from being resolved as X-ray self-shielding is difficult to sustain in a steady-state manner. In multi-dimensional models X-rays are able to scatter around the shielding wind rendering the shielding/shielded wind configuration transient. This contrasts with earlier simulations employing one-dimensional radiative transfer, which produced steady self-shielding structures \citep{proga2000,proga2004,nomura2013,nomura2016,mizumoto2021}. However, this is not the end of the story, as our study relies on several simplifying assumptions. First, we assumed an isotropic source of X-ray radiation. The geometry of the central X-ray source in AGN is still unknown but an anisotropic source where radiation is collimated towards the pole would help in sustaining a wind to higher X-ray level. Second, our wind could develop dense micro-clumps as a result of a radiative instability such as the line-deshadowing instability \citep{owocki1988}. We recently showed that micro-clumping can overcome the over-ionization problem in CVs \citep{mosallanezhad2026}, showing its potential for AGN line-driven winds. Finally, line-driving is not the only mechanism that can produce winds in AGN. Magnetic driving is another candidate and could work in tandem with line-driving.

\section*{Data Availability}
The \textsc{sirocco} and \textsc{pluto} codes used to carry out these simulations are available via the sites \url{https://github.com/sirocco-rt/sirocco} and \url{http://plutocode.ph.unito.it/} respectively. The data files used to generate the figures presented here are available on request.

\section*{Acknowledgements}
NS acknowledges support from the European Research Council (ERC) under the European Union Horizon 2020 research and innovation program (Grant agreement No. 815559 (MHDiscs)). This work was supported by the MHD@Exascale project (reference 22-EXOR-0015) of PEPR Origins (PI: Morbidelli). NS, CK and AM acknowledge support from STFC via grants ST/V001000/1 and UKRI1176. AW was supported by STFC studentship grant 2750006. Partial support for KSL's effort on the project was provided by NASA through grant numbers HST-GO-15984 and HST-GO-16066 from the Space Telescope Science Institute, which is operated by AURA, Inc., under NASA contract NAS 5-26555.  JHM acknowledge funding from a Royal Society University Research Fellowship (URF\textbackslash R1\textbackslash221062).
SAS acknowledges support from STFC grant number ST/X00094X/1. Calculations in this work made use of the Iridis~5 Supercomputer at the University of Southampton. This work used the DiRAC@Durham facility managed by the Institute for Computational Cosmology on behalf of the STFC DiRAC HPC Facility (www.dirac.ac.uk). The equipment was funded by BEIS capital funding via STFC capital grants ST/P002293/1, ST/R002371/1 and ST/S002502/1, Durham University and STFC operations grant ST/R000832/1. DiRAC is part of the National e-Infrastructure. This work was supported by the UK's Science and Technology Facilities Council [ST/M001326/1, ST/P000198/1]. 
We also gratefully acknowledge the use of the following software packages: matplotlib \citep{hunter2007}, Astropy \citep{astropy:2013, astropy:2018, astropy:2022}, {\sc pluto} v4.4 \citep{2007ApJS..170..228M}.

\bibliographystyle{mnras}
\bibliography{biblio}

\begin{appendix}

\section{Isothermal approximation}\label{appendix:isothermal}

\begin{figure*}
    \centering
    \includegraphics[width=180mm]{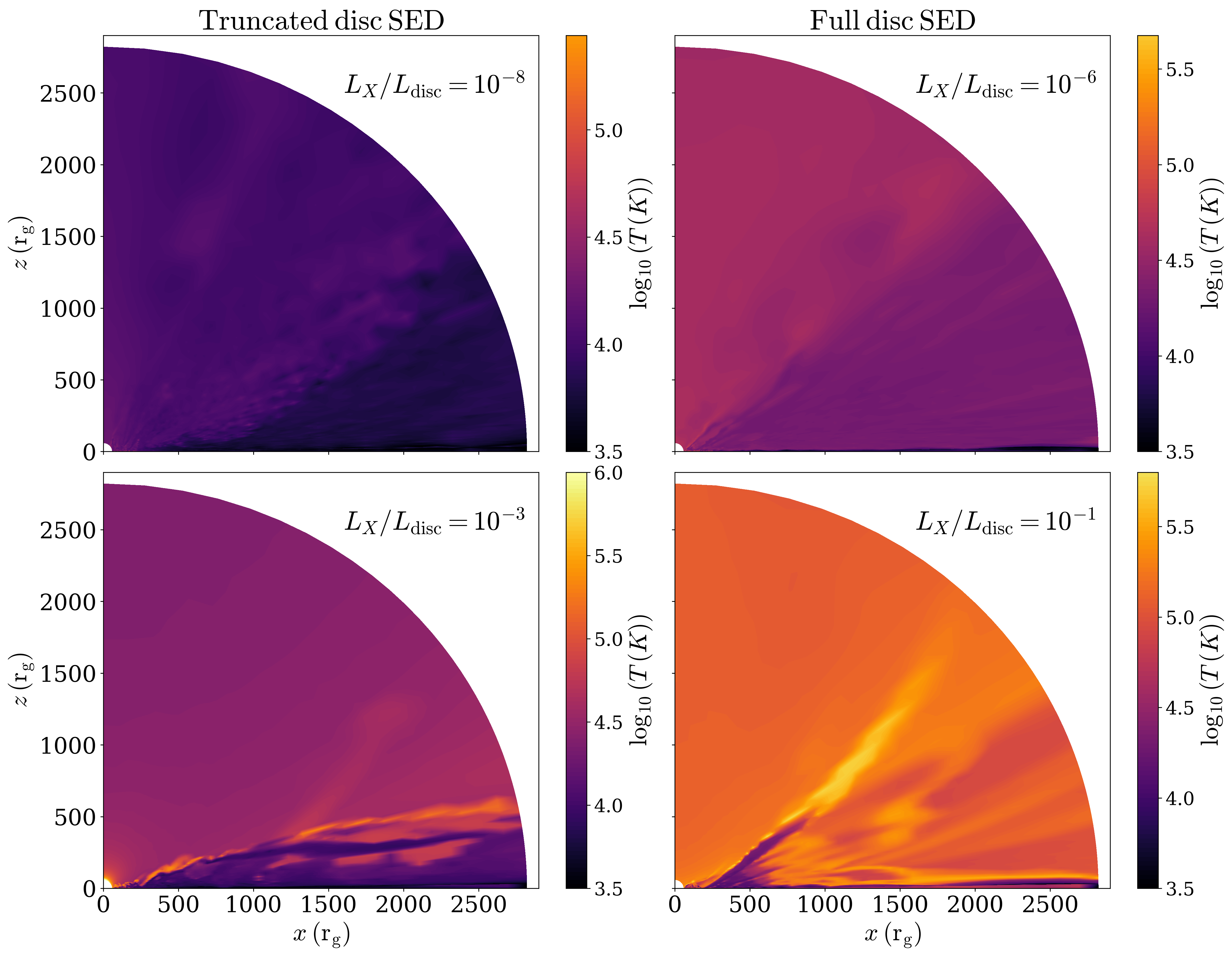}
    \caption{Temperature maps computed from SIROCCO relaxing the isothermal approximation.}
    \label{fig:Tcomputed}
\end{figure*}

As described in \autoref{sec:pluto}, we use an isothermal approximation in PLUTO with a constant temperature of 30,000 K. We check here the impact of this approximation by computing the temperature of the wind from SIROCCO for four simulations, one for each disc SED and one for a low and high X-ray level. We find that the temperature depends on the disc SED and the level of X-rays. For low levels of X-rays, the temperature in the wind is close to the maximum effective temperature of the disc, i.e. close to $10,000$ K for the truncated disc SED and $60,000$ K for the full disc SED so that our isothermal approximation with a temperature of $T=30,000$ K in PLUTO is quite reasonable. As X-rays get more important, we see that they heat the wind over a small layer as they penetrate into it. We also see for the full disc SED at high X-ray level that the X-rays heat even regions that are shielded from direct irradiation from the central source, showing how X-ray scattering is important. Nonetheless, in the densest part of the wind, the temperature is quite similar to that of the low X-ray cases so that our isothermal approximation is likely to be a reasonable first approximation in this case too. Nonetheless, it is possible that we overestimate the presence of the low-ionization plasma necessary for driving in the highest X-ray cases, especially at the beginning of the wind launching when it is the most exposed to X-rays. We will check quantitatively the impact of relaxing the isothermal approximation in a follow-up paper following the method of \cite{mosallanezhad2025}.

\end{appendix}

\label{lastpage}

\bsp	

\end{document}